\documentclass[12pt,preprint]{aastex}

\newcommand{\epseri}{$\epsilon$~Eridani}
\newcommand{\mj}{M$_J$}

\shorttitle{Spitzer/IRAC \epseri Companion Search}
\shortauthors{Marengo et al.}

\begin{document}



\title{A Spitzer/IRAC Search for Substellar Companions \\
  of the Debris Disk Star \epseri}


\author{M. Marengo, S. T. Megeath\altaffilmark{1}, G. G. Fazio}
\affil{Harvard-Smithsonian CfA, 60 Garden St., Cambridge, MA 02138}
\email{mmarengo@cfa.harvard.edu}
\altaffiltext{1}{Ritter Observatory, Dept. of Physics
  and Astronomy, University of Toledo, Toledo, OH 43606}

\author{K. R. Stapelfeldt, M. W. Werner}
\affil{JPL/Caltech, 4800 Oak Grove Drive, Pasadena, CA 91109}

\and

\author{D. E. Backman}
\affil{NASA - Ames, Moffett Field, CA 94035}


\begin{abstract}
We have used the InfraRed Array Camera (IRAC) onboard the Spitzer
Space telescope to search for low mass companions of the nearby debris
disk star \epseri. The star was observed in two epochs 39 days apart,
with different focal plane rotation to allow the 
subtraction of the instrumental Point Spread Function, achieving a
maximum sensitivity of 0.01~MJy/sr at 3.6 and 4.5~\micron, and
0.05~MJy/sr at 5.8 and 8.0~\micron{}. This sensitivity is not
sufficient to directly detect scattered or thermal radiation from the
\epseri{} debris disk. It is however sufficient to allow
the detection of Jovian planets with mass as low as 1~\mj{}
in the IRAC 4.5~\micron{} band. In this band, we detected over 460
sources within the 5.70\arcmin{} field of view of our images. To test
if any of these sources could be a low mass companion to \epseri, we
have compared their colors and magnitudes with models and photometry
of low mass objects. Of the sources detected in at least two
IRAC bands, none fall into the range of mid-IR color and luminosity
expected for cool, 1~Gyr substellar and planetary mass companions of
\epseri, as determined by both models and observations of field M, L
and T dwarf. We identify three new sources which have detections at
4.5~\micron{} only, the lower limit placed on their [3.6]-[4.5] color
consistent with models of planetary mass objects. Their
nature cannot be established with the currently available data and a
new observation at a later epoch will be needed to measure their
proper motion, in order to determine if they are physically associated
to \epseri.
\end{abstract}







\section{Introduction}\label{sec-intro}

\epseri{} (HD~22049, GJ~144, IRAS~03305$-$0937, HIP~16537), along with
Vega, Fomalhaut and $\beta$~Pictoris, is one of the ``Fabulous Four''
nearby main sequence stars found by IRAS to have a cool infrared
excess \citep{gillett86}. This excess is explained as thermal emission
from cool dusty material. A careful analysis of the IRAS data has
resolved the emitting regions for some of these stars, showing a
disk-like structure with a typical size of a few hundreds AU,
analogous to the Kuiper belt of our own solar system
\citep{beichman87}, but with a total mass of dust several orders of
magnitude larger. Given that the timescales for the removal of
primordial material (leftover from the star formation process) is much
shorter than the age of these stars, it is believed that these
structures -- debris disks -- are generated by the collisions between
planetesimals and by cometary activity. As such, debris disks are an
indirect evidence of planetary system formation \citep{backman1993}.

With an age below 1~Gyr (850~Myr according to the most recent
estimate by \citealt{difolco04}, based on fitting VLTI angular
diameter measurements with evolutionary models) and a K2~V
spectral type, \epseri{} is a relatively young star with a mass
slightly lower than the Sun. The proximity of the system, 3.22~pc
distant from the Sun \citep{perryman97}, offers the rare chance to
study in detail the final phases of planetary system formation for a
star similar to the Sun. 

The debris disk of \epseri{} was first imaged at 850~\micron{}
by \citet{greaves98} using the Sub-millimeter
Common-User Bolometer Array (SCUBA) at the James Clerk Maxwell
Telescope in Hawaii. The SCUBA images show a nearly face-on ring
structure, extending as far as 35\arcsec{} from the star, with a maximum
of emission located at a radius of $\sim 18$\arcsec. At the \epseri{}
distance this corresponds to a disk radius of over 100~AU, peaked at the
distance of 60~AU from the star. The nearly face-on orientation of the
disk is in agreement with measured projected rotational velocities ($v
\sin i$) indicating that the star is seen nearly pole-on
\citep{saar97}, with an angle of $i \approx 30\degr \pm 15\degr$ for
the pole of the star (deprojected adopting the rotational period of
11.68~d from \citealt{donahue96}). 

The 1998 sub-millimeter map of \epseri{} has revealed the presence of
a few clumps of emission along the ring. Subsequent maps obtained in
2002 with SCUBA at 850 and 450~\micron{} \citep{greaves05}, and at
350~\micron{} with the SHARC II camera at CSO (Wilner \& Dowell,
personal communication), have confirmed the presence of some of these
clumpy structures. \citet{greaves05} attempted to measure the orbital
motion of the structures detected in their 1998 and 2002 SCUBA
images, and preliminary results suggest a rotation of $\sim 6$\degr{}
counterclockwise \citep{greaves05}. The clumps
have been interpreted as evidences for the presence of a
sub-stellar companion in resonant interaction with the dust: if
confirmed, their measured motion would require an orbital radius of the
resonant body of $\sim 20$~AU. Sophisticated dynamic models of
the sub-millimeter ring suggest an orbital radius for a Jupiter-mass
planet between $\sim 40$~AU \citep{ozernoy00} and $\sim 60$~AU
\citep{quillen02} from \epseri{} (12 -- 20\arcsec{} at \epseri's
distance) to have the required perturbing effect.

Radial velocity measurements have indeed discovered a Jovian class
planet ($M \sin i = 0.86$~\mj\footnotemark[1]\footnotetext[1]{The
  deprojected mass has not been determined yet, but there is an
  ongoing effort by \citet{benedict04} to derive it from the
  astrometric wobble of the star measured with the HST Fine Guidance
  Sensors}) with an estimated semi-major axis of 3.4~AU  and an
eccentricity of 0.6 \citep{hatzes00}, orbiting with a period of
$\approx 7$~yr. This inner planet clearly can not be responsible for
the formation of the structures in the \epseri{} debris disk, as it
would not have any gravitational effect outside the inner $\sim 20$~AU
of the system \citep{moran04}, but it confirms the presence of a
planetary system around the star.

Direct detection of \epseri{} planetary companions has so far
proved unsuccessful. Macintosh et al. (2003, hereafter
\citealt{macintosh03}) carried out near-IR Keck deep Adaptive Optics
(AO) imaging that excluded the presence of
planets with mass larger than 5~\mj{} at spatial scales comparable to
the radius of the dust structures in the debris disk (roughly in a
40\arcsec{} search radius). Proffitt et al. (2004, hereafter
\citet{proffitt04}), while searching for 
disk scattered light in the optical with the Hubble
Space Telescope (HST), compiled a list of $\sim 60$ objects between
12.5\arcsec{} and 58\arcsec{} in a region E from \epseri, none of which,
in absence of a second epoch observation to measure proper motion,
could be identified as a candidate substellar mass companion.

The InfraRed Array Camera (IRAC, \citealt{fazio04}) on-board the
Spitzer Space Telescope \citep{werner04}, thanks to its photometric
system and unprecedented sensitivity in 4 photometric bands at 3.6,
4.5, 5.8 and 8.0~\micron, can uniquely contribute to the search of
substellar mass objects. Brown dwarfs and
Jupiter-size planets are characterized by strong molecular absorptions
features (CH$_4$, NH$_3$, H$_2$O) in the mid-IR wavelength range. The
IRAC passband centered at 3.6~\micron, in particular, is very
sensitive to a deep methane absorption feature at 3.3~\micron, while
the band at 4.5~\micron{} is largely free of molecular opacities. As a
consequence, T-dwarfs and gas giants have nearly unique colors in the
IRAC bands, and can be separated from other classes of astronomical
sources in color-color and color-magnitudes diagrams, based on their
IRAC photometry alone \citep{patten04,patten06}. Based on models by
\citet{burrows03}, a 5~\mj{} planet at the distance of \epseri{} will
have a flux density of 1.3~mJy, and will be easily detected by IRAC
if the artifacts due to the presence of the bright star can be
mitigated. 

To take advantage of these unique characteristics of IRAC, we
carried out deep imaging of \epseri{} in all IRAC bands, as part of
the ``Fabulous Four debris disk star'' Spitzer Guaranteed Time
Observation (GTO) program (Program ID 90). This program consists of
the observation of Vega, Fomalhaut, $\beta$~Pictoris and \epseri{} 
with all Spitzer instruments, in order to investigate the structure of
their debris disks and to search for substellar companions. Results
concerning the structure of the disks of Fomalhaut and Vega as seen by
the Multiband Imaging Photometer for Spitzer (MIPS, \citealt{rieke04})
and the InfraRed Spectrograph (IRS, \citealt{houck04}) have already
been published \citep{stapelfeldt04, su05}. A paper based on IRS
spectroscopy and MIPS imaging of \epseri{} disk is in preparation
(Backman et al. 2006). We present here the result of the search for
substellar companions around \epseri{} made with IRAC. In
Section~\ref{sec-obs} we describe the IRAC observations and the data
reduction, with particular emphasis on the technique adopted to remove
the \epseri{} starlight diffracted on the focal plane, to allow
detection of faint sources as close as possible to the star. In
Section~\ref{sec-limits} we discuss our detection limits for disk
extended emission and for low-mass companions. In
Section~\ref{sec-search} we describe the results of our companion
search and compare them with previous searches. Finally, we present
our conclusions in Section~\ref{sec-conclude}.


\section{Data acquisition and reduction}\label{sec-obs}

Due to its proximity to the Sun, \epseri{} is a very bright star in
the mid-IR, with a 2MASS K magnitude of 1.76 \citep{cutri03}. Faint
companions and diffuse circumstellar emission can be detected only
after a careful subtraction of the bright central point source. The
Spitzer Space telescope is specially suited for this task, thanks to
the exceptional stability of its optics and pointing system, that
allows a precise measurement and reproducibility of its instrumental
Point Spread Function (PSF). The IRAC observations for the whole
``Fabulous Four'' program were designed to maximize the ability of
subtracting the stellar PSF.

Each star was observed in two epochs, typically one month apart, to
take advantage of the different roll angle of the spacecraft, which is
determined by the relative position of the spacecraft with the
Sun. This strategy allows to position the main features of the PSF
(the diffraction spikes created by the tripod supporting the secondary
mirror and a few artifacts created by the electronics), which are
fixed with respect to the detector pixel grid, in different positions
on the sky. This allows a complete coverage of the sky around each of
the stars.

\epseri{} was observed on January 9, 2004 (AOR 4876032) and on February
17, 2004 (AOR 4876288). The roll angle offset between the two epochs
is of 20.15\degr, clockwise. Each observation consisted in a sequence
of 12~s Full Array frames (10.4~s integration time), dithered on the
IRAC arrays on a 36 Position Reuleaux triangle, using the small dither
scale, with each position repeated 9 times. The total integration
time was 3379.6~s (over 56 minutes) on-source, for each IRAC band.
The source was set at the center of the array,
where the PSF is cleaner from artifacts.
The other ``Fabulous Four'' stars were
observed with the same technique, and with a total integration time
scaled according to the brightness of the source. A standard star
($\epsilon$~Indi) was also observed, with the same total integration
time as \epseri, to contribute to the measurement of the instrumental
PSF. The total integration time was estimated on the basis of the
sensitivity required to detect a Jupiter mass planet in the vicinity
of \epseri. The individual frame exposure time (10.4~s) was chosen
because it allows to get a long total integration time divided in a
manageable number of individual frames, while still avoiding excessive
saturation of the star on the detector. Shorter frame times are
available, but could not be used because it would have been very
inefficient to reach the required total integration time and the large
field of view necessary to search for widely separated low mass 
companions. The total field of view imaged in each band was 5.78\arcmin,
slightly larger than the IRAC field of view of 5.21\arcmin, due
to the dithering pattern. 

Basic data reduction and calibration was done with the Spitzer Science
Center (SSC) pipeline, version S10.0.1 (first epoch) and S10.5.0
(second epoch). The IRAC pipeline returns the individual exposures
calibrated in physical units of surface brightness. This calibration is
valid only for point sources, as it was obtained by matching the flux
of standard stars modeled by \citet{cohen03} within an aperture of
12.2\arcsec. Note that point
source photometry with different sets of apertures and sky annuli and
the photometry of extended sources requires the use of appropriate
aperture corrections. These aperture corrections and the other
parameters used in the absolute photometric calibration (including the
``FLUXCONV'' factors used to convert the raw data in DN/s into the
science calibrated data in MJy/sr) are listed in Table~\ref{tab-cal}.

For each epoch, separately, we have created a mosaic combining the
individual frames, using the SSC Mosaicer MOPEX on a final grid with a
pixel scale of 0.4\arcsec/pix (a factor 3 smaller than the original
IRAC pixel scale, of $\approx$1.22\arcsec/pix) to leverage the high
coverage for a better sampling of the PSF. We have chosen to maintain
the same orientation in the final mosaic as the individual frames, to
have the PSF features oriented with the same angle in the mosaics of
the two epochs. Cosmic rays and other outliers have been removed using
the MOPEX temporal outlier module. This procedure was repeated
identically for all ``Fabulous Four'' sources, including the standard
star, in all epochs.


\subsection{PSF subtraction}\label{ssec-psf}

\begin{figure}
\begin{center}
\plotone{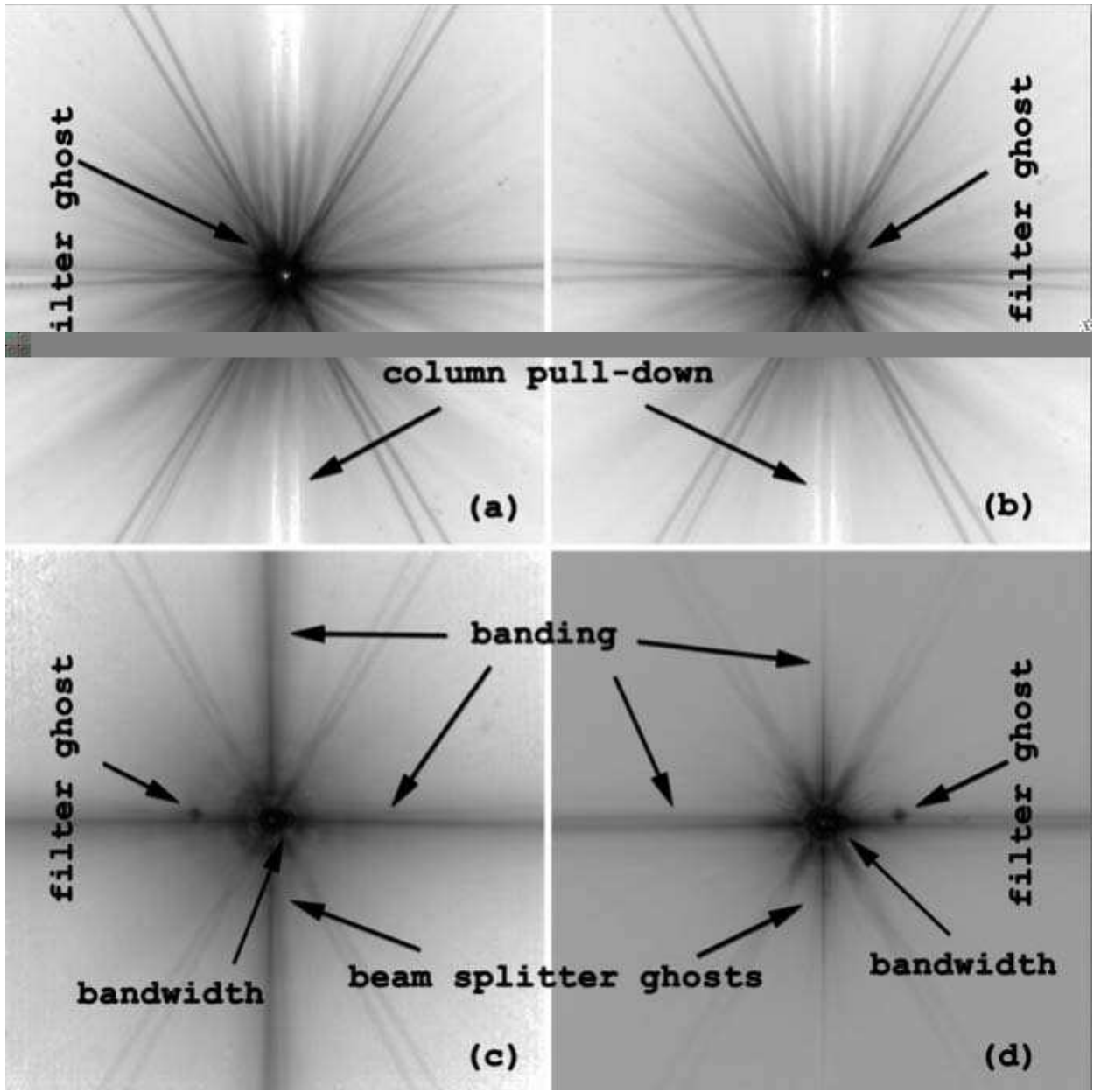}
\caption{The IRAC PSF at 3.6~\micron{} (a), 4.5~\micron{} (b),
  5.8~\micron{} (c) and 8.0~\micron{} (d). The PSF is shown in
  inverted logarithmic color scale, from zero intensity (white) to
  saturation level (black). The size of each panel is the same as the
  IRAC field of view ($\sim 312$\arcsec). The main artifacts are
  marked. The ``beam splitter ghosts'' at 5.8 and 8.0~\micron{} are
  the faint ``kinks'' visible along the vertical banding where the
  arrows point. Note that there is column pull-down also at 5.8 and
  8.0~\micron, which is however not visible due to the predominant
  effect of the banding. Muxbleed is also present at 3.6 and
  4.5~\micron, and can be seen after PSF subtraction in
  Figure~\ref{fig-dropout}. Note that the horizontal spikes at 5.8 and
  8.0~\micron{} are hidden below the banding.}\label{fig-psf} 
\end{center}
\end{figure}

Figure~\ref{fig-psf} shows the ``Fabulous Four'' program PSF we have
derived by combining the mosaics of Vega (2 epochs) and
$\epsilon$~Indi (1 epoch). Note that
Vega itself is a star surrounded by a debris disk, but the infrared
excess in IRAC bands is too small to be detectable by IRAC 
\citep{su05}. Vega can thus be considered as a point source for
the purpose of building our PSF. The three mosaics have been shifted
by matching the position of the unsaturated diffraction spikes,
rescaled according to the relative brightness of Vega and
$\epsilon$~Indi and then coadded. The final PSF is clean of
background stars, that have been $\sigma$-clipped in the coadding
phase. 

A special version of the PSFs we have generated for this project, 
that combines all the stars of the ``Fabulous Four'' program with the
exclusion of $\beta$~Pictoris (whose debris disk is indeed extended at
IRAC wavelengths), is available at the SSC web
site\footnote{http://ssc.spitzer.caltech.edu/irac/psf.html}
for different pixel
scales. These public PSFs also include an unsaturated core derived
from the observation of faint standard stars as part of the IRAC
photometric calibration project. A paper describing the construction
of these PSFs and their main characteristics is in preparation
(Marengo et al. 2006). 

It is clear from Figure~\ref{fig-psf} that the IRAC PSF has a lot of
structures and artifacts that are the main limiting factor for
observing faint emission around bright sources. These artifacts cover
the whole area of the IRAC arrays, and can be divided in two
categories. The 6 diffraction spikes (each of them composed by two
initially braided and then diverging individual spikes), the smooth
extended PSF tails, the diffraction rings and the PSF ``ghosts''
(``filter ghosts'', small ``ring-like'' and ``cross-like'' structures
at the left of the PSF peak at 3.6 and 5.8~\micron{} and the right at 4.5 and
8.0~\micron{}, and ``beam splitter ghosts'', fainter structures below
the 5.8 and 8.0~\micron{} PSF peaks) are part of the ``optical
PSF''. These structures are 
linear in intensity with the fluency of the source (product of its
surface brightness with the integration time) and can thus be easily
subtracted even with a PSF derived from stars of different brightness than
the source, as long as the PSF stars are observed with the same dither
pattern and in the same position on the IRAC detectors (the position
of the ghosts and the orientation of the diffraction spikes does
depend on the array position).  The negative columns (``column pull-down'',
white area in Figure~\ref{fig-psf}) and the positive fencing pattern
(``muxbleed'', visible in Figure~\ref{fig-dropout} after PSF
subtraction) in the 3.6 and 4.5~\micron{} bands, the large positive 
``crosses'' (vertical and horizontal ``banding'') and the saturated
multiple ``echos'' on the right of the saturated core (``bandwidth effect'')
in the 5.8 and 8.0~\micron{} bands are instead ``electronic
artifacts'' generated by the detector and the readout electronics and
artifacts resulting from internal scattering of IR light within the
detectors. The
column pull-down also affects the 5.8 and 8.0~\micron{} bands, but can
be hidden below the stronger banding. The electronic artifacts are
not linear with the source fluency and can only be subtracted using a
PSF obtained from stars with the same signal as the source. 

The PSF that we have derived from Vega and
$\epsilon$~Indi is only effective for subtracting the optical
structures of the PSF. To eliminate the electronic artifacts it is
necessary to use an alternative approach, based on the observation
of \epseri{} in two epochs, with two different roll angles. By
using the second epoch as a PSF model for the first epoch (a PSF with
exactly the same fluency of the source), we can effectively subtract
the optical PSF \emph{and} the electronic artifacts, while preserving the
background sources that are rotated in the two epochs with respect to
the PSF. In a two epoch subtracted frame the background sources from
one epoch will be positive, and the ones from the other epoch will be
negative. The trade-off for this technique is that for a very crowded
field there is the risk of aliasing ``positive'' with ``negative''
sources when they superpose by chance, due to the rotation of the
field.

To have the best possible coverage of the sources in the \epseri{}
field, we have 
applied both PSF subtraction techniques in all IRAC bands. For each
band we have thus produced 4 PSF-subtracted frames: (1) first epoch
subtracted with the PSF; (2) second epoch subtracted with the PSF; (3)
first epoch subtracted with the second epoch and (4) second epoch
subtracted with the first epoch (3 and 4 are of course the same,
except for the orientation on the sky and the sign).

\begin{figure}
\begin{center}
\plotone{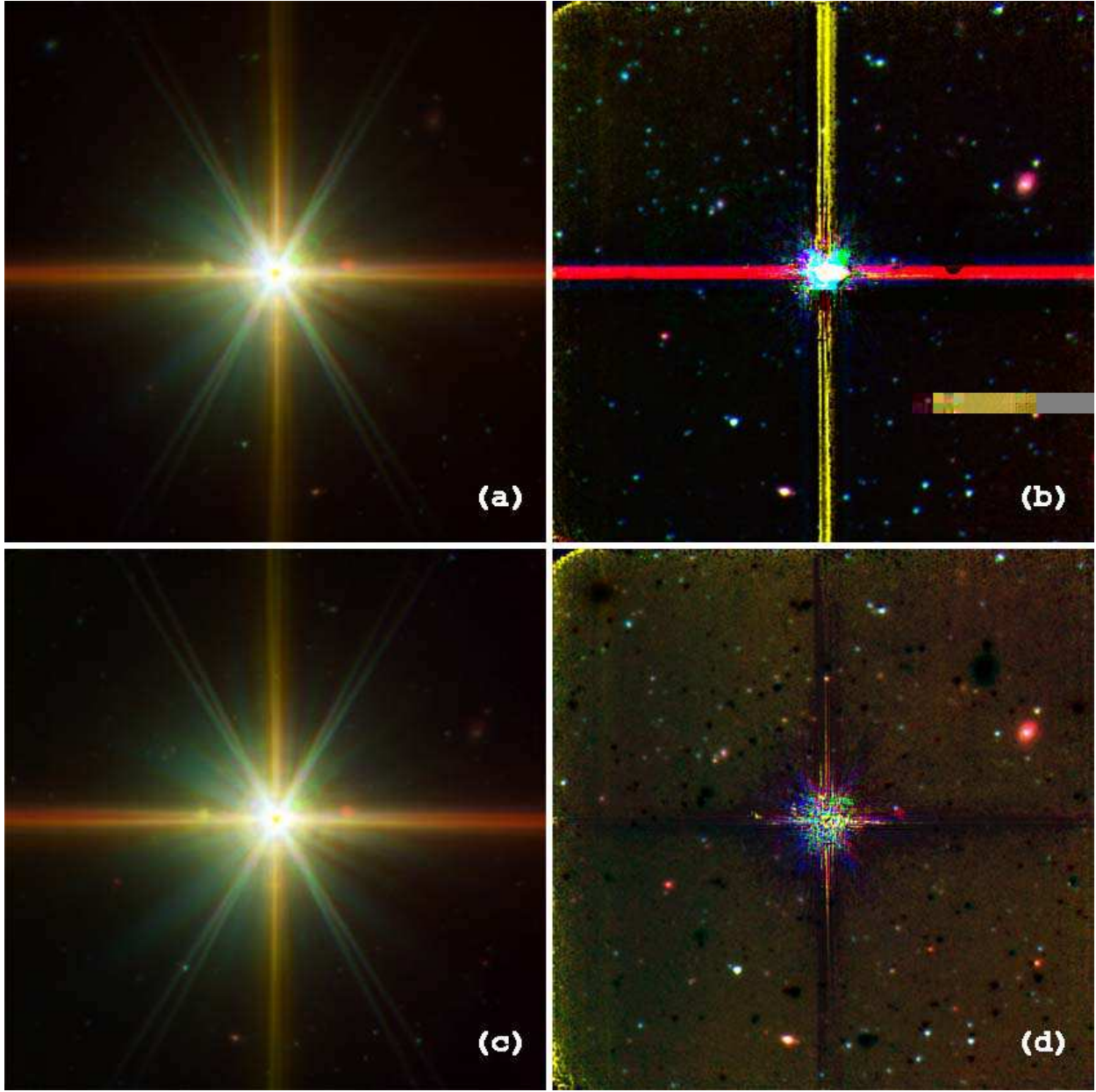}
\caption{\epseri{} before (left) and after (right) PSF
  subtraction. In the color composite, blue is 3.6~\micron, green is
  4.5~\micron, yellow is 5.8~\micron{} and red is 8.0~\micron. Note
  the rotation of the field of view between epoch 1 (a) and epoch 2
  (c). Panel (b) is epoch 2 subtracted with the PSF constructed from
  Vega and $\epsilon$~Indi. Panel (d) is epoch 2 subtracted with epoch
  1. The left panels are shown in logarithmic color scale, while the
  PSF-subtracted images are shown in linear scale from -0.5~MJy/sr to
  1.0~MJy/sr. The total field of view is
  334\arcsec.}\label{fig-psfsub}
\end{center}
\end{figure}

Figure~\ref{fig-psfsub} shows an example of the two PSF subtraction
techniques (right panels) compared with the original images before PSF
subtraction (left panels). The top right panel shows the epoch 2 image
subtracted with the PSF: note the remaining ``cross'' due to imperfect
subtraction of the ``electronic'' artifacts. The bottom right panel
shows the second epoch image subtracted with the first: note the
better subtraction of the ``electronic'' artifacts, but the occurrence
of aliasing for some of the field sources. Both PSF subtraction methods
are able to remove most of the \epseri{} light scattered by the PSF,
and bring forth the numerous fainter sources in the field. Limits to
this technique due to residual noise after the PSF subtraction are
described in the following section.

The PSF subtraction technique described here allows to measure the
brightness of \epseri{} with considerable precision, by determining
the scaling factor between the \epseri{} image and the PSF (which is
normalized, by construction, as the calibrated image of Vega).
This scaling factor is determined by the fitting
routine better to than 1\%, which translates into an accuracy of
$\sim$0.01~mags. The measured magnitudes, in each epoch, are listed in 
Table~\ref{tab-mags} (the
conversion from magnitudes to fluxes uses the IRAC Vega fluxes also
reported in the \citealt{irac04}). Note that there is no photometric
variation between the two epochs, with the exception of $\sim 2.7$\%
change in the 5.8~\micron{} flux (which is however within 3$\sigma$
of the estimated photometric error).


\section{Sensitivity limits for extended and point sources}\label{sec-limits}

The PSF subtraction methods described in the previous section are
able to remove most of the \epseri{} light scattered on the
IRAC arrays by the Spitzer optics and the detector electronics. The
subtraction, however, leaves a number of artifacts and residual noise
that limits our capability to search for faint companions and extended
structures. These residuals are especially strong in close proximity
to the star, precluding the search for planetary companions in the
inner \epseri{} system.

\begin{figure}
\begin{center}
\plotone{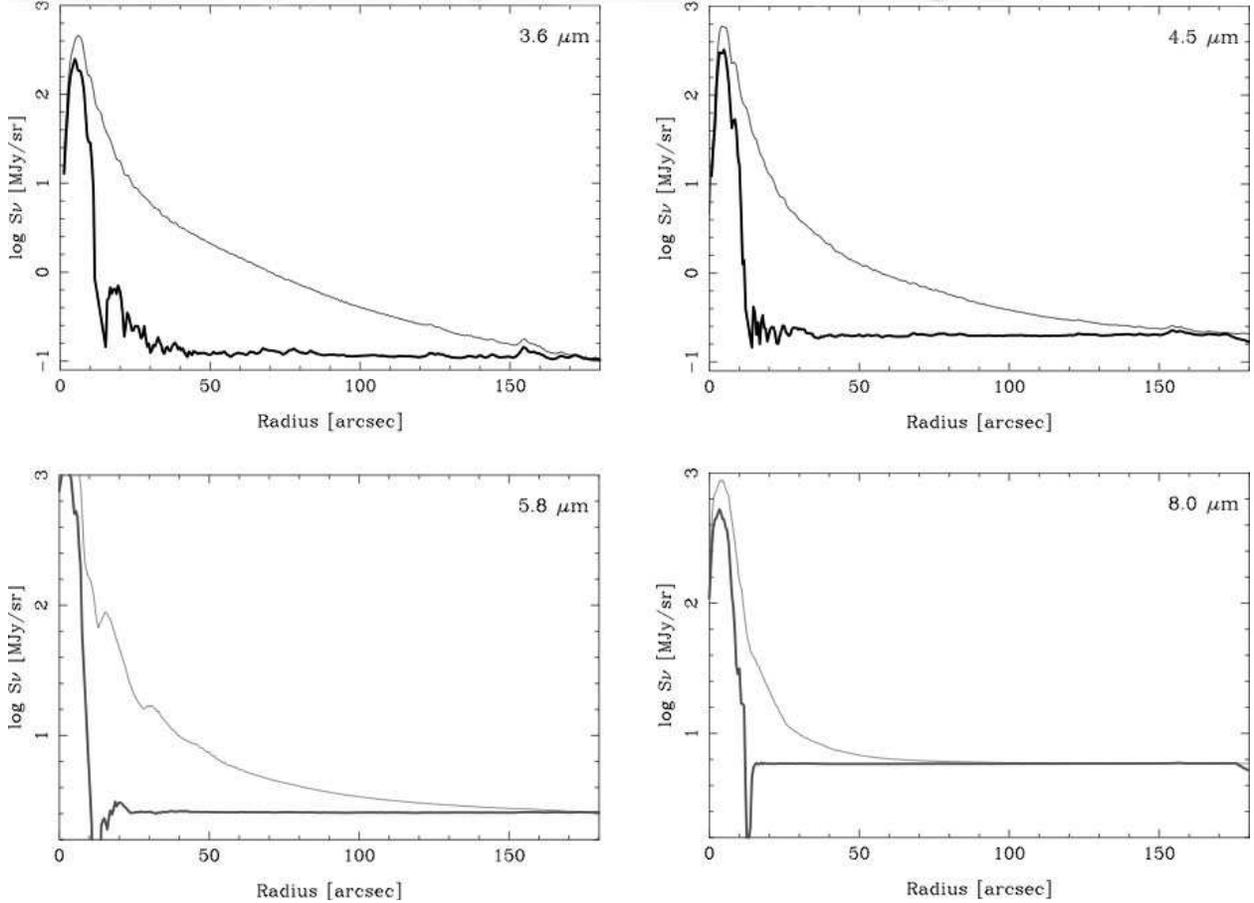}
\caption{\epseri{} (epoch 2) radial profiles before (thin line) and
  after (thick line) subtraction with the IRAC PSF in the four IRAC
  bands. The profile is a circular average within a wedge of 40\degr{}
  width, with a position angle of 120\degr{} from the vertical,
  clockwise (to avoid diffraction spikes and other artifacts). The
  background pedestal level has not been subtracted from the profiles
  presented in this figure (even though it is removed from the PSF
  subtracted images used in the rest of the
  analysis).}\label{fig-profiles}
\end{center}
\end{figure}

Figure~\ref{fig-profiles} shows the radial profiles of \epseri{}
before (thin line) and after (thick line) PSF subtraction, along a
wedge with Position Angle 120\degr{} and width 40\degr{} (positioned
to avoid diffraction spikes and electronic artifacts). Note the
presence of a background pedestal, in part due to an imperfect
subtraction of the dark frames, and in part due to an electronic
pedestal produced when a bright source is incident on the detector.
This background pedestal is particularly strong in IRAC
images containing bright sources, as it is the case with \epseri, and
is larger at longer wavelengths. The pedestal level in our
observations was $\sim 0.1$, 0.2, 2.5 and 6.0~MJy/sr at 3.6, 4.5,
5.8 and 8.0~\micron{} respectively, and was successfully removed from
the PSF-subtracted images (but not from the profiles shown in
Figure~\ref{fig-profiles}) during PSF subtraction by adding a constant
value to the scaled PSF, even though a small residual slope was left
in the final 5.8~\micron{} images. 

The IRAC arrays become severely non linear when the signal exceeds
the $\sim 30,000$ -- 50,000~DN range (see Table~\ref{tab-cal}).
Above this limits the
image is rapidly saturated. This effect can be observed in the radial
profiles as a surface brightness depression in the inner 5-7\arcsec{}
(16-22~AU), where all signal is lost, making the core of the star
looking like a ``donut''. Outside this radius the PSF profiles rapidly
fall, even though the residuals after PSF subtractions are still high
within $\sim 14$\arcsec{} from the star ($\sim 45$~AU). These
residuals are mainly due to the high photon noise of the bright
source, the buildup of strong latents during the long total
integration, and to the intrinsic limitations in the PSF construction
method: even small changes in the dither pattern, combined with the
undersampling of the IRAC PSF (especially severe at 3.6 and 4.5
microns), can cause small but significant deviations in the PSF
profiles, that result in elevated residual noise where the PSF signal
is stronger. Outside a $\sim 14$\arcsec{} radius the PSF subtraction
is however very effective to lower the PSF signal to the photon noise
level.


\subsection{Limits on the debris disk detection}\label{ssec-diskd}

\citet{proffitt04} have used the STIS CCD camera on the Hubble Space
Telescope to search for an optical counterpart of the submillimeter
disk in scattered light. The result of their search was
inconclusive, but they set an upper limit of 25 mag/arcsec$^2$ for
the optical surface brightness of the dust at the distance of 55~AU
from the star (17\arcsec, roughly the distance where the millimetric
disk emission peaks), within the ``clear aperture'' wide spectral
bandpass of the STIS camera (from 0.2 to 1.02~\micron). At the nominal
wavelength of 0.7~\micron, this limit corresponds to a surface
brightness of 0.011~MJy/sr.

Given that the spectral energy distribution of a K star like \epseri{}
peaks in the near-IR, the intensity of the light scattered from the
disk rapidly decreases at wavelengths longer than 1~\micron, at least
until the thermal emission of small grains starts to be significant
(for $\lambda \ga 10$~\micron). This means that at IRAC wavelengths
the total disk surface brightness will be very
small. \citet{proffitt04} have developed a model of the disk surface
brightness based on a previous model by \citet{li03}, fitting the
850~\micron{} and IRAS emission and including the contribution from
scattered optical light. This model, that assumes the presence of
highly porous particles, and a rather flat grain size distribution
from 1~\micron{} to 1~cm, predicts that the surface brightness of the
disk at 3.6~\micron{} and at the distance of 55~AU would be less than
0.002~MJy/sr, and even lower in the other IRAC bands (see
Figure~11 in \citealt{proffitt04}). This surface brightness is too
small to be detected in our PSF-subtracted images.

Figure~\ref{fig-rms} shows the sensitivity of our images for extended
sources, after subtraction of the \epseri{} starlight. We initially
computed the \emph{RMS} noise in a running boxcar of 4\arcsec{} size
for the PSF and two-epochs subtracted images, obtaining 4 maps for each
IRAC band. The 4 maps differ in that the residuals of the PSF
subtraction, and thus the residual noise, depend on the PSF
subtraction method and on the rotation of the field with respect to
the array coordinates. Since we
are interested in the best case sensitivity achievable combining both
epochs and both PSF subtraction methods, we have constructed a final
map by taking the minimum \emph{RMS} value, in each pixel, of the
individual maps. We have then corrected the minimum \emph{RMS} maps
for the ``infinite aperture correction'' necessary to recalibrate the
maps for the case of extended sources (listed in Table~\ref{tab-cal}).

The contours show that the residual \emph{RMS} noise increases closer
to the star, as expected, and along the main PSF artifacts
(diffraction spikes and electronic artifacts). The large oval shape at
the left from the star at 4.5~\micron{} is enhanced noise due to a pupil
ghost. At the distance of 55~AU from the star the sensitivity is greater
than 1~MJy/sr in all channels, making impossible to test the accuracy
of the \citet{proffitt04} model. Even as far as 1\arcmin{} from the
star, the maximum sensitivity (0.01~MJy/sr at 3.6 and 4.5~\micron{}
and 0.05~MJy/sr at 5.8 and 8.0~\micron) is still not sufficient to
detect the scattered light from the disk.

The \citet{proffitt04} model predicts a total flux of 0.12, 0.14, 0.19
and 0.40~mJy at 3.6, 4.5, 5.8 and 8.0~\micron. \citet{sheret04}, using
a simple model of a thin ring with solid silicate and organic grains
with a size distribution similar to the interstellar medium
\citep{mathis77}, but ranging from 1.75~\micron{} to 5~m, found a
larger flux of 1.7, 1.5, 0.97 and 0.53~mJy in the IRAC bands. In both
cases the predicted fluxes are much smaller than the accuracy of our
IRAC photometry (see Table~\ref{tab-mags}), and the expected infrared
excess from the disk over the photospheric emission cannot be detected
by our measurement.


\subsection{Limits on low mass companions detection}\label{ssec-planetd}

Models shows that, at least in the case of a faint debris disk like
the one around \epseri, the detection of scattered light from the
circumstellar dust is beyond the sensitivity limits of our
PSF-subtracted images. The main goal of IRAC imaging, however, is the
search for faint companions with sub-stellar mass. From
the profiles shown in Figure~\ref{fig-profiles} it is clear that we
have no point source sensitivity closer than $\sim 14$\arcsec{}
(45~AU) from the star, because of saturation, nonlinearity and high
photon noise preventing usable PSF subtraction very close to the
star. This rules out the possibility to search for the radial velocity
planet at 3.2~AU, which would be well within the saturated area in our
images. A different story, however, concerns the possibility to
detect the planet whose presence at a distance between 40 and 60~AU
(12\arcsec{} -- 20\arcsec) is predicted by some dynamic models of the
observed clumps in the debris disk. Even though such planet has been 
unsuccessfully searched by \citet{macintosh03}, the superior
sensitivity of IRAC and the photometric system more suited for the
observation of low mass objects may give us the chance of detecting
this body, if it exists and if its mass is above our detection
limits. The wide field of view of the IRAC maps (covering an area with
size over 1,200~AU around the star) also allows us to search for widely
separated planets that may have been ejected from the system, due to
planet-planet interactions \citep{ford01} in case the total
protoplanetary disk mass was not sufficient to dampen the
eccentricity of planets forming in the system.

Since the discovery of Brown Dwarfs \citep{becklin88,nakajima95},
several groups have started to compute detailed model atmospheres of
substellar mass objects to predict what would be their emission
spectra and luminosity. More recently, these models became very
sophisticated, to include large lists of molecular and alkali
opacities, dust formation and settling, improved treatment of line
pressure broadening and the effects of water clouds and ice formation,
enabling them to compute reliable spectra extending in the IR
domain. Figure~\ref{fig-burrows} shows a family of such models,
computed by \citet{burrows03} for 1~Gyr old substellar objects with
mass from 1 to 25~\mj. The model spectra are shown in the wavelength
range covered by the IRAC photometric system, and have been rescaled
for the distance of \epseri. As mentioned in
section~\ref{sec-intro}, the spectrum of substellar objects in the
IRAC bands is modulated by deep CH$_4$ and H$_2$O absorption
features. Methane absorption is particularly strong in the
3.6~\micron{} IRAC band, where it can depress the flux by a few
orders of magnitude below the continuum, with increasing strength
with decreasing temperature (and thus decreasing mass, for a fixed age).
A second broad methane feature is present within the
8.0~\micron{} band, while water absorption affects the flux around
5.8~\micron. The 4.5~\micron{} IRAC band, however, is relatively free
of molecular absorption, and thus offers the best chances of detection
for molecular rich, cool substellar objects.

We have estimated the magnitude and colors of the atmosphere models
presented in \citet{burrows03} in the IRAC photometric system by
integrating each model with the IRAC bandpasses, normalized by the flux
of Vega:

\begin{equation}
m_{(band)} = -2.5 \, \log \left[ \frac{\int_{band} F_\nu(\lambda) \,
    R_\nu(\lambda) \, \lambda \, d\lambda}{\int_{band}
    F^{(Vega)}_\nu(\lambda) \, R_\nu(\lambda)\, \lambda \, d\lambda}
  \right] 
\end{equation}

\noindent
where $F_\nu(\lambda)$ is the model spectrum, $F^{(Vega)}_\nu(\lambda)$
is the model spectrum of Vega, derived by Robert Kurucz for Martin
Cohen (this is the same model used for the absolute photometric
calibration of IRAC), and $R_\nu(\lambda)$ is the IRAC Spectral
Response Curve from the IRAC Spitzer Science Center web
site\footnote{http://ssc.spitzer.caltech.edu/irac/spectral\_response.html}
(August 9, 2004 version).

\begin{figure}
\begin{center}
\plotone{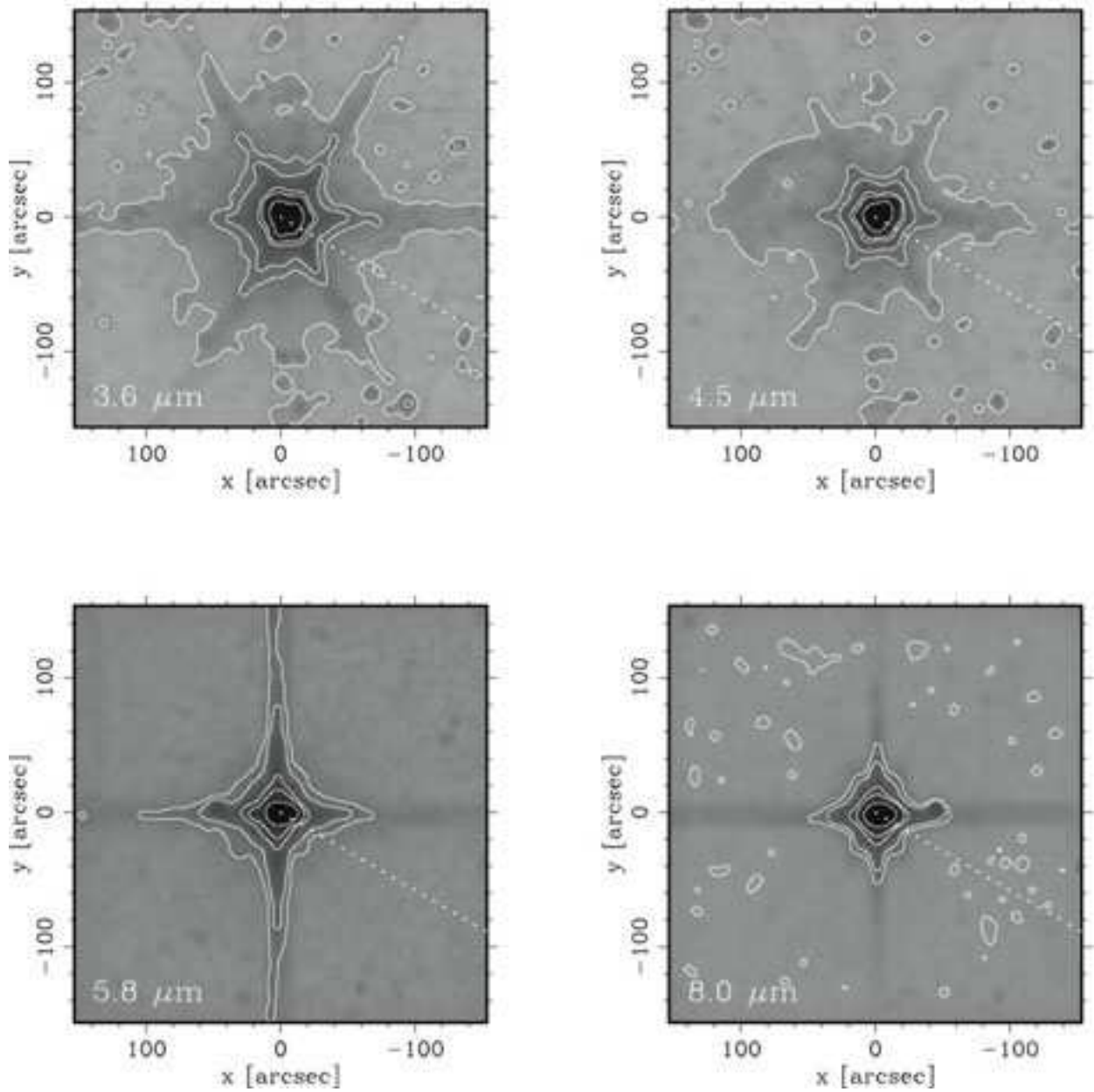}
\caption{Sensitivity maps, in array coordinates, computed as the RMS
  noise of the PSF-subtracted images, estimated in a running boxcar of
  size 4\arcsec{}. The contours show the RMS noise of 5, 1, 0.5, 0.1,
  0.05 and 0.01 MJy/sr, with the highest noise level at the
  center. The dotted line indicates the direction of the cut used to
  plot the radial profiles in Figure~\ref{fig-profiles} and
  \ref{fig-mags}.}\label{fig-rms}
\end{center}
\end{figure}

\begin{figure}
\begin{center}
\plotone{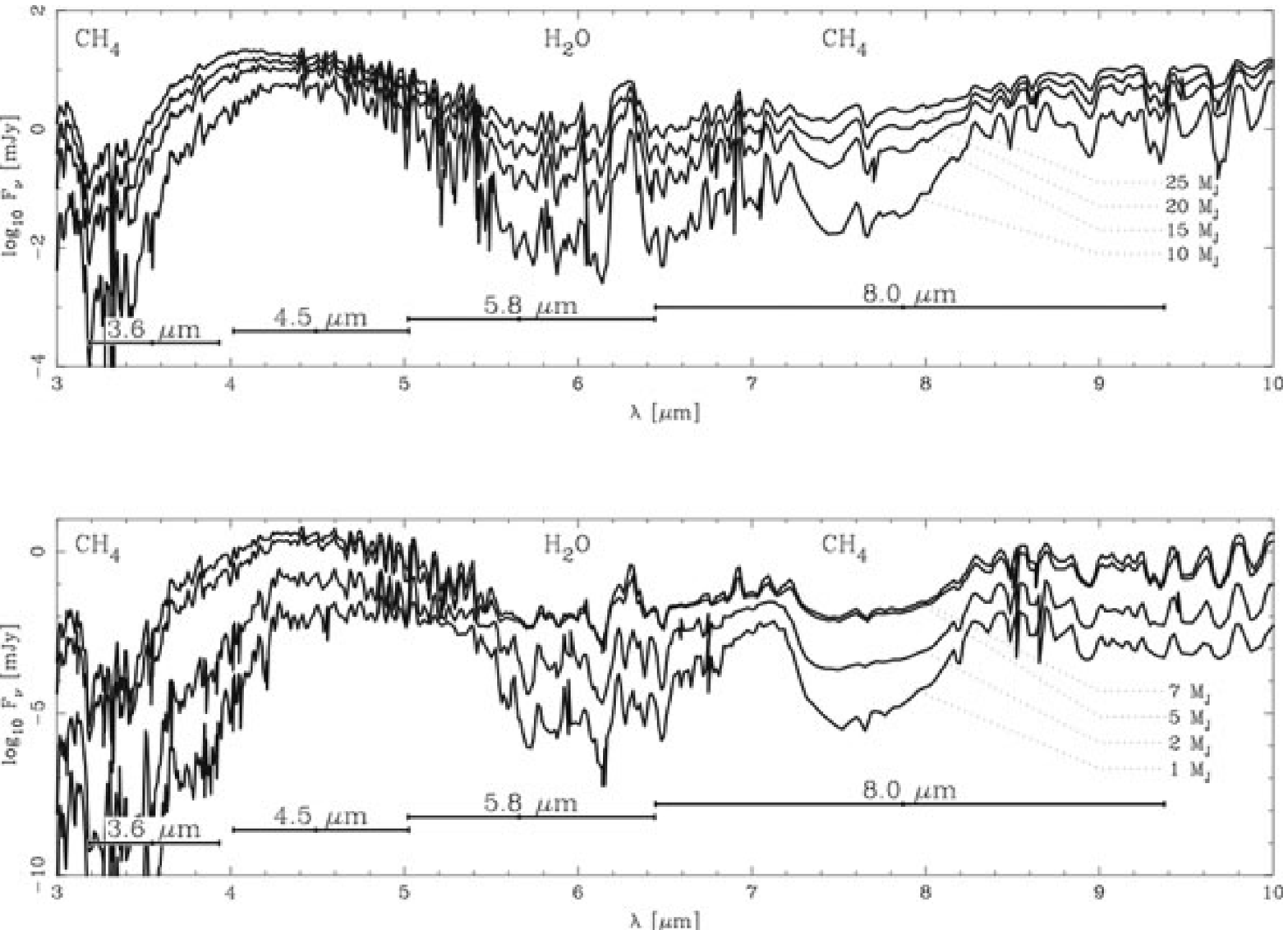}
\caption{Model spectra of gas giants and Brown Dwarfs from
  \citet{burrows03}, computed for 1, 2, 5, 7, 10, 15, 20 and 25
  \mj{} 1~Gyr bodies. The predicted flux density has been rescaled for the
  distance of \epseri{}. The main spectral features in the IRAC bands
  are produced by molecular absorption of CH$_4$ (in the 3.6~\micron{}
  and 8.0~\micron{} bands) and H$_2$O (in the 5.8~\micron{} band). The
  band at 4.5~\micron{} is relatively free of absorption, and is where
  Jupiter size planets would appear brighter.}\label{fig-burrows}
\end{center}
\end{figure}

\begin{figure}
\begin{center}
\plotone{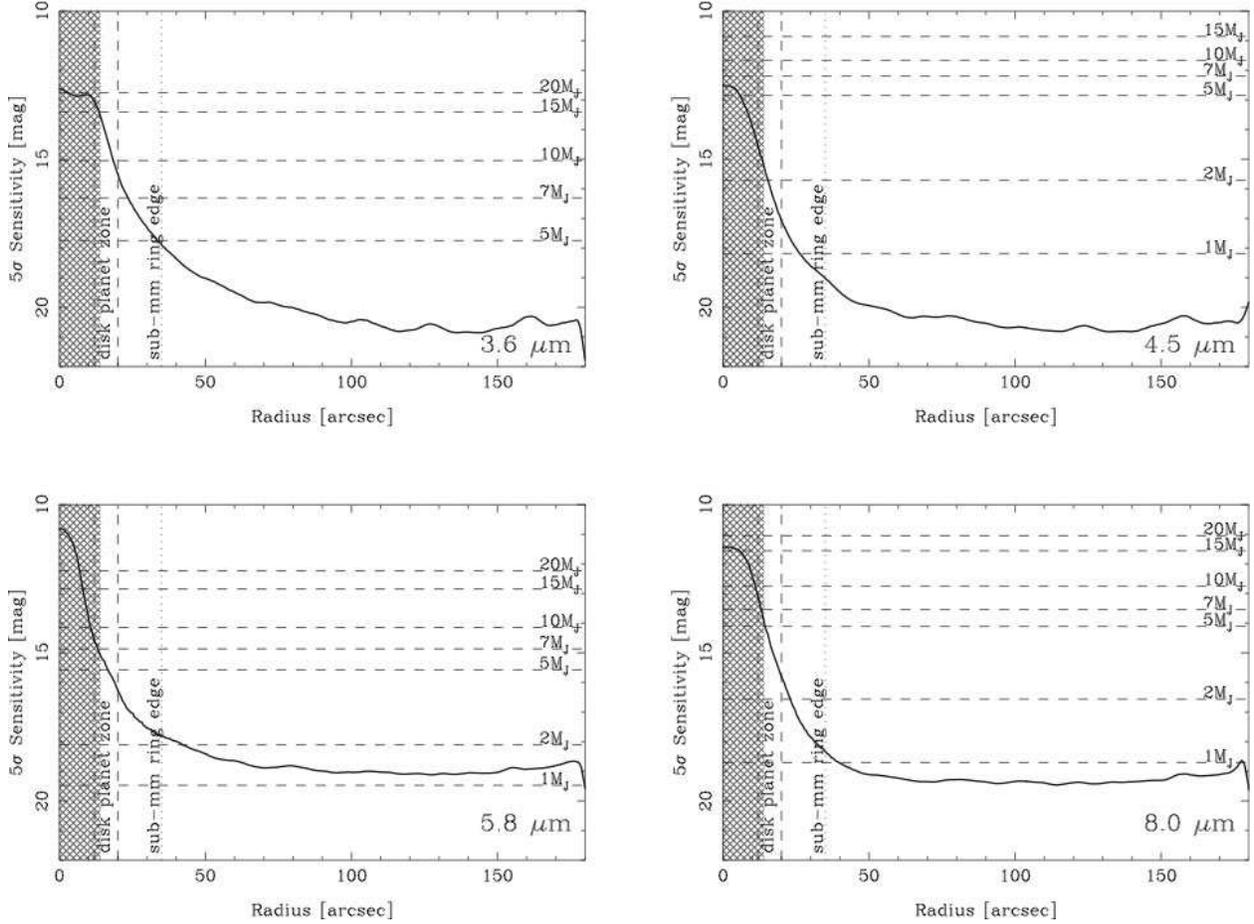}
\caption{Detection limits for giant planets and Brown Dwarfs based on
  models computed by \citet{burrows03}. The solid line shows the
  5$\sigma$ point source sensitivity (within an aperture of
  2.4\arcsec{} radius) derived from the RMS maps in
  Figure~\ref{fig-rms}, along a cut with position angle 120\degr{}
  and 40\degr{} width. The
  cross-hatched region within a radius of 14\arcsec (45~A.U.) from the
  star should be excluded because of saturation and high residual
  noise after PSF subtraction. The dashed vertical lines show the
  region where dynamic models predict the presence of a resonant body
  (between 40 and 60~A.U.), while the dotted vertical line marks the
  outer radius of the sub-millimetric ring at 35\arcsec{}
  (112~A.U.). }\label{fig-mags}
\end{center}
\end{figure}

The predicted magnitudes are shown in Figure~\ref{fig-mags}, and
compared with the $5 \sigma$ point source sensitivity of our PSF-subtracted
images along the same radial cut used in Figure~\ref{fig-profiles}. We
have derived the point source sensitivity from the maps in
Figure~\ref{fig-rms} by converting the \emph{RMS} noise into
\emph{Noise Equivalent Flux Density} ($NEFD)$ inside an aperture of 2
IRAC pixels radius (2.4\arcsec):

\begin{equation}
m^{(5 \sigma)}(x,y) = -2.5 \, \log \left[ \frac{5 \cdot
    NEFD(x,y)}{F_{(Vega)}} \right]
\end{equation}

\noindent
and

\begin{equation}
NEFD(x,y) = 2.3504 \cdot 10^{-5} \cdot r_A \cdot \sqrt{\pi}
(pixscale)^2 \cdot acorr(r_A) \cdot RMS(x,y)
\end{equation}

\noindent
where $m^{(5 \sigma)}$ is the point source sensitivity in magnitudes
at the $(x,y)$ coordinates in the PSF-subtracted images, $RMS(x,y)$ is
the sensitivity map shown in Figure~\ref{fig-rms}, $F_{(Vega)}$ is the
flux of Vega in the IRAC bandpasses reported in the \citet{irac04},
and the other quantities are geometrical factors necessary to convert
surface brightness into flux, and include the
aperture radius $r_A$, the pixel scale of the PSF-subtracted images
(0.4\arcsec/pix) and the aperture correction $acorr(r_A)$ for 2 IRAC pixel
apertures, also from the \citet{irac04}, listed in Table~\ref{tab-cal}.

Figure~\ref{fig-mags} shows that indeed the IRAC band
most sensitive to detect cool companions
is the one at 4.5~\micron. Outside the area where the image is
contaminated by saturation, non-linearities and photon noise from the
central bright star (cross-hatched in the
figure), our PSF-subtracted images are sensitive to Jupiter mass 
planets. Just inside the inner edge of the ring
we are sensitive to planets with a mass of $\simeq 1$~\mj,
and below 1~\mj{} in the area outside the
debris disk. The strong methane absorption in the 3.6~\micron{} band
increases considerably the minimum detectable mass at this wavelength,
so much that in the region inside the disk where dynamic models
suggest the presence of planets gravitationally perturbing the disk,
we are only sensitive to bodies with mass larger than 7~\mj. The
sensitivity in the two remaining IRAC bands (5.8 and 8.0~\micron) is
intermediate between these two extreme cases. Note that, with the
exception of the 3.6~\micron{} band, outside a radius of 45~AU 
our PSF-subtracted images are at least a factor 2 more sensitive in
terms of mass than
the Keck K band images in \citet{macintosh03}, and at least 5 times
more sensitive outside the submillimeter ring.


\section{Companion search}\label{sec-search}

Even though \epseri{} is relatively far from the galactic plane, the
field around the star shows a large number of background sources. The
challenge is determining if any of them can be a gravitationally bound
low mass companion, instead of a faint background star or
galaxy, or an artifact from the PSF subtraction. The similar search
made by \citet{proffitt04} in the optical concluded that most of the 
detection were in fact background
galaxies. Given the sensitivity of IRAC to red extragalactic objects,
we should expect a similar result.

As mentioned above, the colors of substellar objects in the IRAC
photometric system are very peculiar. The colors of brown dwarfs and
giant planets at the distance of \epseri{} can be estimated from
\citet{burrows03} models, as described in section~\ref{ssec-planetd}.
For all background sources having good photometry
in at least two or three IRAC bands, it is thus possible to
investigate their nature on the basis of color-color and
color-magnitude diagrams. Planets below 5~\mj, however, may be
detected in only one IRAC channel, at 4.5~\micron, because of the
lower sensitivity at 5.8 and 8.0~\micron, and the deep methane
absorption feature depressing their 3.6~\micron{} flux below our
sensitivity level. Given that our two epochs are too close in time to
allow for proper motion measurements, for these ``3.6~\micron{} dropout''
sources  a new observation at a later epoch is needed to check if they
are comoving with the \epseri{} system.


\subsection{Source detection and photometry}\label{ssec-phot}

The PSF subtraction procedure leaves behind a large number of
point-source-like artifacts that can easily confuse the detection
algorithms available in most photometry packages. In particular,
analysis based on the shape of the point sources by Gaussian or PSF
fitting of all structures above a certain threshold would reject a
large number of valid sources whose shape has been distorted by the
proximity with the PSF artifacts, or by a nearby ``aliasing''
source in the two epoch subtracted frames. Given our requirement to
be inclusive, not to miss the proverbial ``needle in the hay stack''
that would be a planet orbiting \epseri, we have relaxed our
requirements starting with a list including \emph{any} detection with S/N
$\ga 5$ within an aperture of 2.4\arcsec{} radius.
We have then selected the real from the spurious sources by
visually inspecting each detection, and by comparing the two epochs
and the results from the two different PSF subtraction methods
described in section~\ref{ssec-psf}. With this procedure we have
obtained a list of 467 sources with magnitudes between 12 and 20,
reliably detected in at least one epoch at 4.5~\micron{} within the
IRAC field of view.

\citet{fazio04b} shows that, outside the plane of the Galaxy, most of
the sources detected by IRAC with a magnitude $\ga 13$ at 4.5~\micron{} are
background galaxies. Table~1 in that paper shows that the total number
of background galaxies with 4.5~\micron{} magnitudes between 15 and 20
within the area of our maps is $\sim 920$. This is consistent with
our number of detections, taking into account the large area at the
center of our PSF-subtracted images where our sensitivity limits is
well below the 20th magnitude because of the high PSF subtraction residuals.

For source detection and photometry we have used the package
``PHOTVIS'' developed by Rob Gutermuth \citep{gutermuth04}, based on
the ``APER'' aperture photometry package part of the IDL ``ASTROLIB''
library. The size of the aperture was set to 2.4\arcsec, with a sky
annulus with inner and outer radii of 2.4\arcsec{} and 7.2\arcsec. To
convert the mosaic data (in MJy/sr surface brightness units) into
magnitudes we have derived zero point magnitudes based on the Vega
fluxes adopted for the camera's absolute calibration, listed in
Table~\ref{tab-cal}. 

The uncertainty of each photometric measurement was
estimated as the sum in quadrature of the scatter in the sky values, the
uncertainty in the mean sky brightness and the random photon noise on
source (estimated from the source counts converted in electrons). Note
that this error analysis tends to underestimate systematic errors due
to imperfect subtractions of the PSF and PSF artifacts that
alter the sky level and introduce non-Gaussian noise.

We have then merged the individual source lists in each band and PSF
subtraction method. In each IRAC band we have averaged the individual
photometry in the two epochs and PSF subtractions, weighted on their
S/N ratio. Typical uncertainties of the final photometry
are $\sim 0.05$~mag at 3.6 and 4.5~\micron, and $\sim 0.1$~mag at 5.8
and 8.0~\micron. Sources close to \epseri{} or along PSF subtraction
artifacts may have a larger uncertainty in one of the two epochs, and
depending on the PSF subtraction method. We have checked the
consistency of the photometry of sources detected in both epochs, and
found that it is generally within the error margin, except when a PSF
artifact or source aliasing is present. In these case,
we assume an uncertainty of $\sim 0.2$~mag, representative of the
maximum spread observed in the photometry of sources detected in both
epochs in at least two IRAC bands.


\subsection{Color-color and color-magnitude diagrams}\label{ssec-knn}

\begin{figure}
\begin{center}
\epsscale{0.75}
\plotone{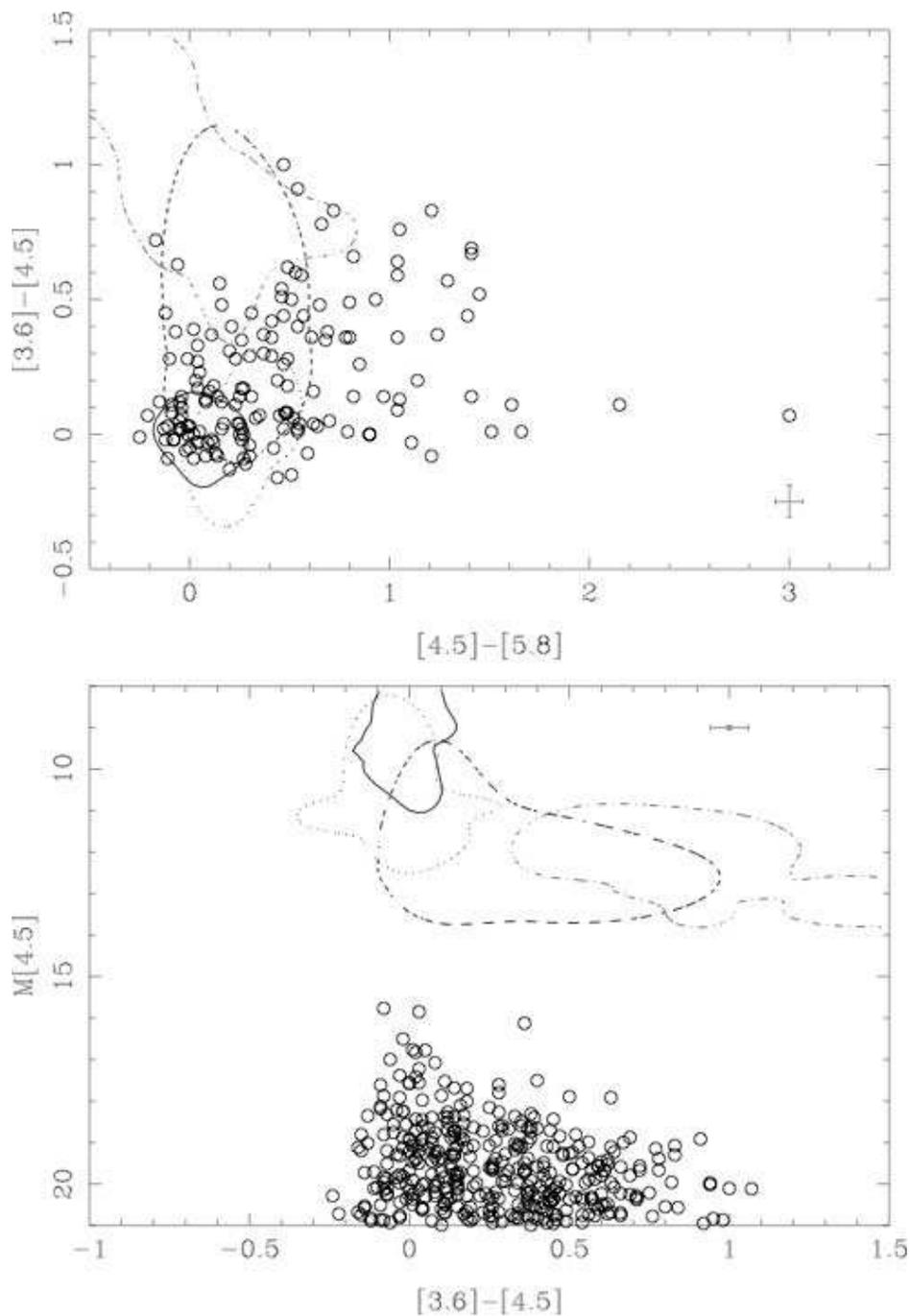}
\caption{Color-color and color-magnitude diagrams of all
  point sources detected in the field around \epseri. The contours
  are the region selected with the $k$NN method for M dwarfs (solid
  line), L dwarf (dotted line), early T dwarfs (T$<$5, dashed line)
  and late T dwarfs (T$\ga$5, dot-dashed line). Note that many sources
  have the right colors to be M, L or T dwarfs, but none of them has
  the correct brightness to be an M, L or T companion of
  \epseri, according to the colors and magnitudes of field M, L and T
  dwarfs from \citet{patten06}.}\label{fig-knn-MLT} 
\end{center}
\end{figure}

The top panel of Figure~\ref{fig-knn-MLT} shows the [3.6]-[4.5]
vs. [4.5]-[5.8] color-color diagram of the sources detected in all
three bands. Note that many of the sources are clumped around zero
colors, indicating that they are regular background stars. A
significant number of sources, however, have one or both colors red:
this is expected for extragalactic sources having significant PAH
emission in the longer wavelength IRAC bands, or red-shifted galaxies
\citep{huang04,barmby04}. The bottom panel shows the [3.6]-[4.5] color
vs. the 4.5~\micron{} absolute magnitude that the sources detected in
both 3.6 and 4.5~\micron{} bands would have if they were at the same
distance (3.22~pc) of \epseri.

The contours plotted on the figure enclose the locations where sources
with the colors of field M, L and T dwarfs are located in the diagram. These
contours have been derived from data collected as part of the IRAC
Guaranteed Time observations of M, L and T dwarfs \citep{patten04,
  patten06}, using a variation of the weighted $k$-Nearest Neighbor
Method ($k$NN, \citealt{fix51}) adapted for astronomical
applications. The method, commonly used in diagnostic medical
studies is a multi-variate analysis very effective in the automated
classification of items in different groups. In this work we have used
$k = 3$ (3rd nearest neighbor) and a Gaussian kernel to bias the
metric towards the 1st nearest neighbor. Sources inside each region
differ from the class templates (the M, L, and T field stars observed
by \citealt{patten06}) by less than 1$\sigma$ error in photometry or
color. A detailed description of this method will be given in a
separate paper (Marengo \& Sanchez 2006, in preparation). 

The top panel of Figure~\ref{fig-knn-MLT} shows that many of the
sources detected around \epseri{} have colors compatible with the
colors observed for field M, L and T dwarfs \citep{patten06}. The
bottom panel of the same figure, however, shows that all the sources
detected at least at 3.6 and 4.5~\micron{} are much fainter than all
the field M, L and T dwarfs in the \citet{patten06} sample. Although
the observed field M, L and T dwarfs have a range of masses and ages
which may not be appropriate for \epseri, lower mass brown dwarfs and
planets at a given [3.6]-[4.5] color cannot be much fainter than the
$M_{4.5}$ range shown for the field objects. The reason is that the
radii of $\sim 1$~Gyr brown dwarfs and planets span a relatively small
range of values, and are relatively constant independent of age and
mass \citep{burrows97,chabrier97}. Hence their luminosity depends
primarily on temperature, and consequently color.

\begin{figure}
\begin{center}
\epsscale{0.75}
\plotone{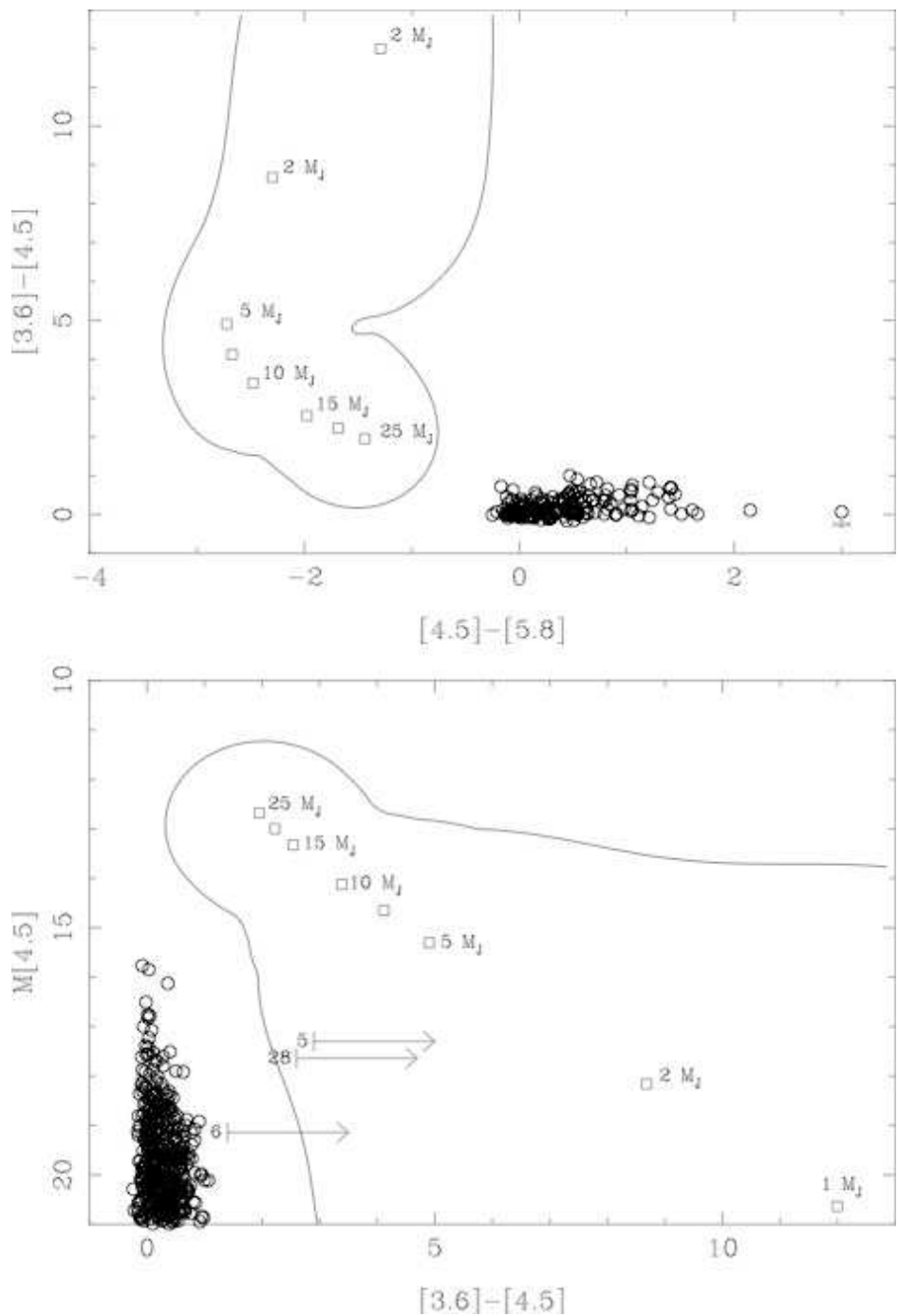}
\caption{Color-color and color-magnitude diagrams of 850~Myr giant planet
  model tracks from \citet{burrows03}, compared with the sources
  detected in the \epseri{} field. Square points denote the colors and
  magnitudes of model planets with mass from 25 to 1~\mj. The solid
  line encloses the region in the diagrams where candidate planets may
  be found, according to our $k$NN method. None of the sources detected
  at both 3.6 and 4.5~\micron{} has the correct colors and
  magnitudes predicted by \citet{burrows03} models. The arrows
  indicate the photometry and colors of the three 3.6~\micron{}
  ``dropout sources'' listed in
  Table~\ref{tab-dropout}.}\label{fig-knn-burrows}
\end{center}
\end{figure}

\begin{figure}
\begin{center}
\epsscale{0.90}
\plotone{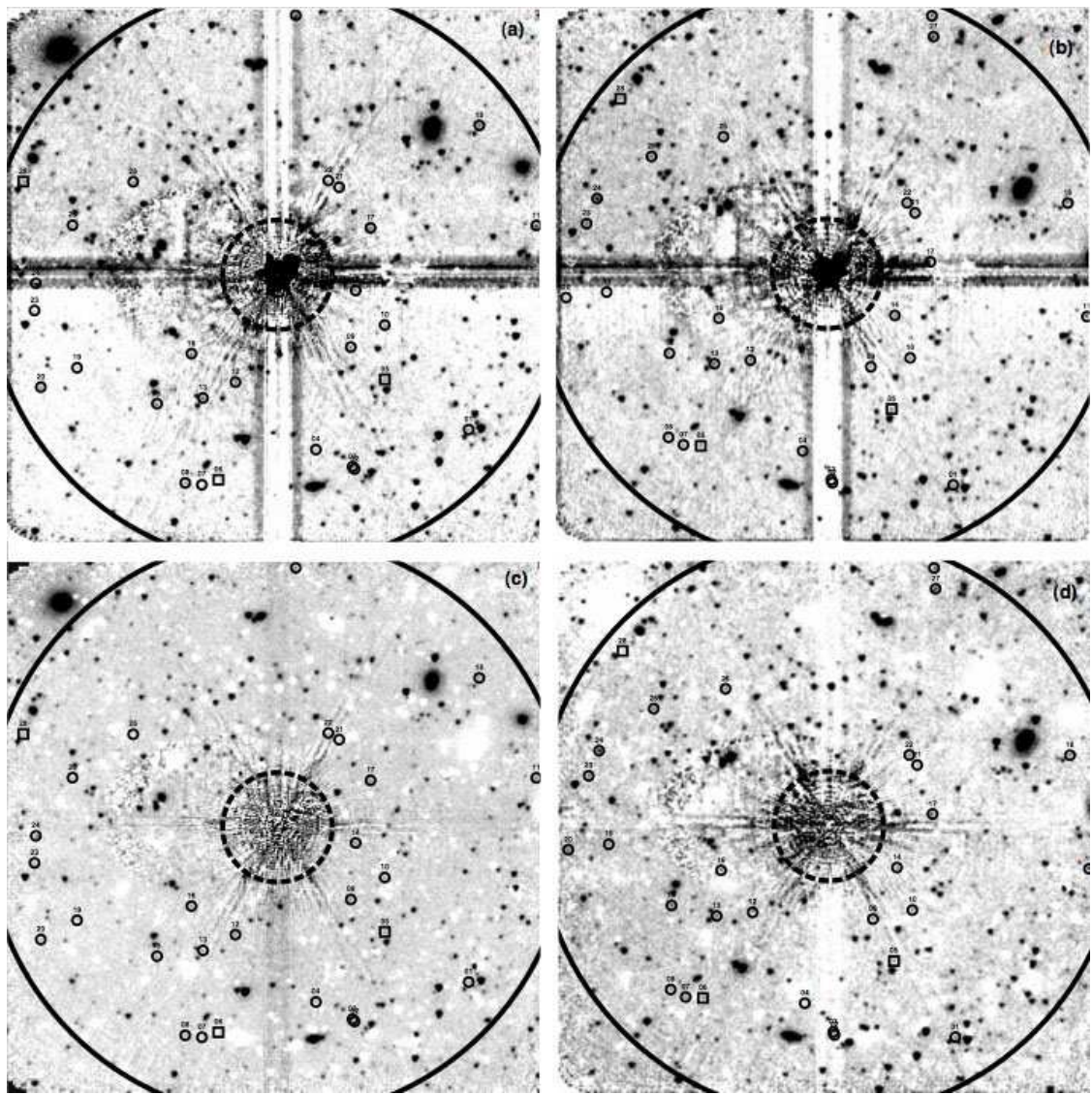}
\caption{PSF-subtracted IRAC 4.5~\micron{} images of \epseri{}
  background field: (a) epoch 1 subtracted with the PSF, (b) epoch 2
  subtracted with the PSF, (c) epoch 1 subtracted with epoch 2 and (d)
  epoch 2 subtracted with epoch 1. The figure uses an inverted linear
  color scale from background level to 0.1~MJy/sr. The inner circle
  marks the maximum radius of the sub-millimeter ring (112~AU,
  corresponding to 35\arcsec{} at the \epseri{} distance). The outer
  circle marks the distance of 600~AU. Circle points mark the
  location of point sources for which only a marginal detection ($\la
  5 \sigma$) at 3.6~\micron{} is available. The three sources with no
  detection at all at 3.6~\micron{} are indicated with square
  points.}\label{fig-dropout}
\end{center}
\end{figure}

This is demonstrated in Figure~\ref{fig-knn-burrows} by the
\citet{burrows03} models of 1~Gyr old brown dwarfs and giant
planets. The large area covered by the $k$NN contours is determined by
the large systematic error we have assumed for the models,
approximately as large as the color separation between the models of
individual mass. This is a prudent assumption, in our requirement to
be inclusive, given that these models have never been directly tested
towards real extrasolar planets. Even so, none of the detected sources
possess the right combination of colors and magnitudes to be a
candidate planet or T dwarf, as all the planet models have a
[3.6]-[4.5] color much redder, and a [4.5]-[5.8] color much bluer (due
to H$_2$O absorption at 5.8~\micron) than the sources detected in at
least two IRAC bands.


\subsection{The 3.6~\micron{} dropout sources}\label{ssec-dropout}

Figure~\ref{fig-knn-burrows} shows that a planet with 5~\mj{} would
have a 4.5~\micron{} absolute magnitude of $\sim 15.3$ and a [3.6]-[4.5]
color of 4.9. This means that the 3.6~\micron{} magnitude of that
planet at the distance of \epseri{} would be 20.2, which is below our 
detection limit. Such a planet can be detected in the 4.5~\micron{}
images, but would drop out from our 3.6~\micron{} images. Bodies with
mass below $\la 5$~\mj{} cannot be found on the basis of their
color with the $k$NN method described in the previous section, and
their identification should rely on proper motion measurements.

The time lag between our two epochs, however, is too short to yield a
measurable shift of \epseri{} and any companion. The proper motion of
\epseri{} is of -0.977\arcsec/yr in R.A. and 0.018\arcsec/yr in
Dec \citep{perryman97}, which means that during the 39 days
between the two epochs the star moved by $\sim 0.1$\arcsec. This is a
factor 2 less than our best case centroiding accuracy ($\sim
0.2$\arcsec{} at 4.5~\micron, where the IRAC pixel scale has the
optimal PSF sampling). Comparison with other catalogs is
not possible, with the exception of the few sources in common with the
\citet{proffitt04} and \citet{macintosh03} surveys, as
described in the next section.

Table~\ref{tab-dropout} lists all the sources, within a radius of
600~AU (approximately the area covered by our images) that have a
5$\sigma$ detection at 4.5~\micron, but that are not detected at the
5$\sigma$ level at 3.6~\micron. Most of these sources do have,
however, a marginal 3.6~\micron{} detection, and their colors (even
allowing for the large uncertainties in their 3.6~\micron{}
photometry) suggest them to be red background galaxies. Only three
sources are completely undetected at 3.6~\micron. Their 4.5~\micron{}
fluxes and [3.6]-[4.5] colors (estimated from the 3.6~\micron{}
sensitivity limit in Figure~\ref{fig-rms}) are compatible with the
colors and magnitudes of planets with masses $\la 3$~\mj{} (sources
5 and 28 in Table~\ref{tab-dropout}) and $\la 2$~\mj{} (source 6 in
Table~\ref{tab-dropout}), but not with faint red-shifted
galaxies (which typically have [3.6]-[4.5] $\la 0.5$). Huang et
al. (personal communication) measured that out of $7.3 \cdot 10^4$ sources
detected by IRAC in an area of 2\arcdeg{} $\times$ 10\arcmin{} with a 3
hours exposure only 6 have a 4.5~\micron{} magnitude between 17.5 and
20, and [3.6]-[4.5] $>$ 1.5. It is thus very unlikely to have 3 such
sources within a single IRAC field of view. A careful inspection of
the individual 4.5~\micron{} PSF-subtracted images however suggests
these sources to be multiple (and possibly extended, at least in the
case of source 5), which may indicate that their colors may have been
altered by chance superposition with higher than average PSF
subtraction artifacts. The real nature of these three detections cannot
be determined with the current data, and a new observation to confirm
that they are real, and to measure their proper motion, is necessary to
determine if they are physically associated to \epseri.


\subsection{Comparison with Keck AO and HST searches}\label{ssec-macintosh}

\begin{figure}
\begin{center}
\epsscale{0.90}
\plotone{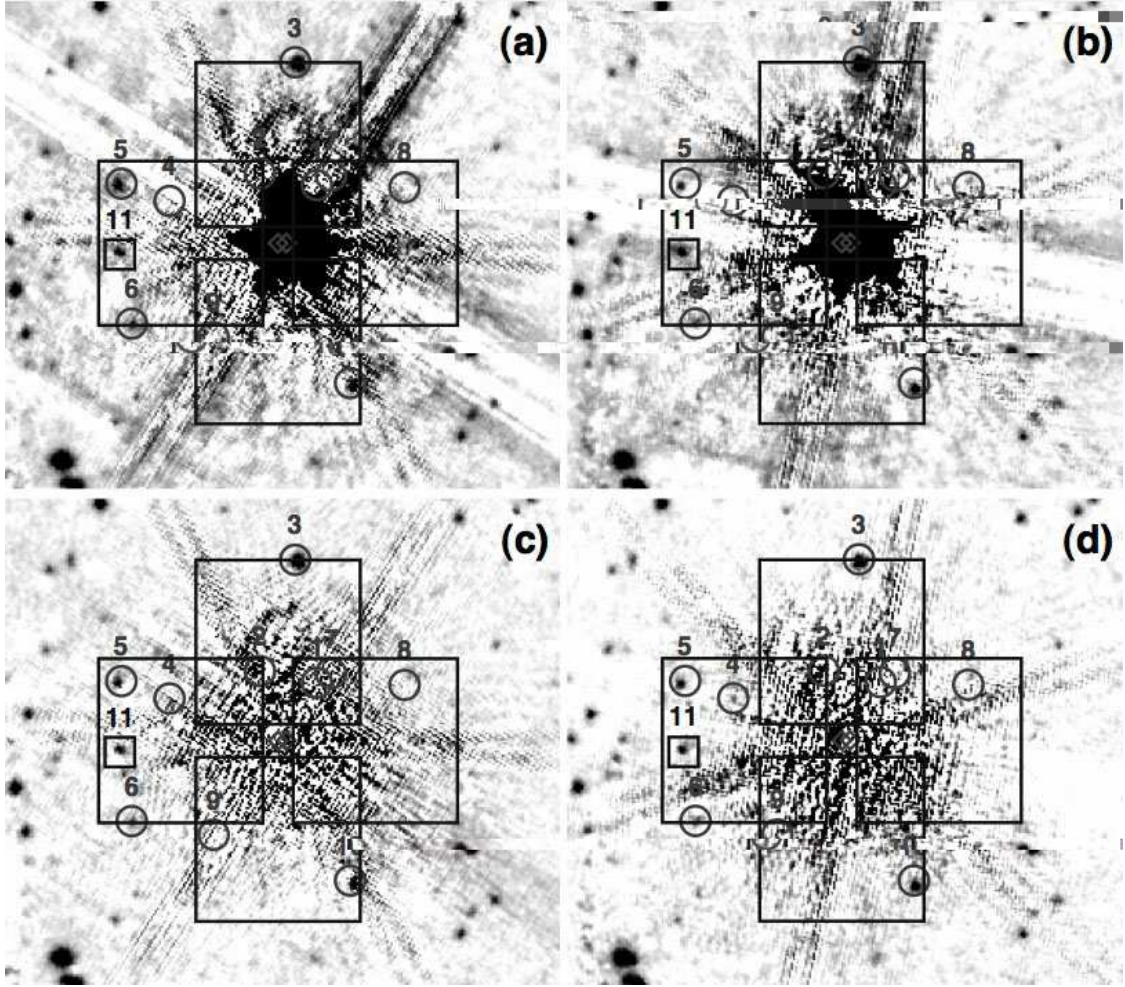}
\caption{IRAC 4.5~\micron{} images of the area surrounding \epseri{}
  that was searched by \citet{macintosh03}: (a) epoch 1 PSF
  subtracted; (b) epoch 2 PSF subtracted; (c) epoch 1 - epoch 2; (d)
  epoch 2 - epoch 1. The large squares are the field of view of the
  individual Keck images (40\arcsec{} size). The circles 
  marks the positions of the sources found in the \citet{macintosh03}
  search extrapolated to their expected position at the epoch of the
  IRAC observations. The small box indicates a source found in our
  IRAC images that was missed in \citet{macintosh03}. The two little
  diamonds indicate the position of \epseri{} at the time of the first
  Keck epoch (left diamond), and the IRAC (right diamond)
  observations.}\label{fig-macintosh} 
\end{center}
\end{figure}

Even though none of our detected sources (with the possible exception
of the three 3.6~\micron{} ``dropout sources'' described in the
previous section), do not have the colors or magnitudes to be low mass
companions of \epseri, it is however useful to compare the result of
our survey with the previous \citet{macintosh03} search.
Figure~\ref{fig-macintosh} shows the location, in our 4.5~\micron{}
images, of the 10 objects identified by \citet{macintosh03} in their K
band Keck AO maps within an area roughly $\approx 40$\arcsec{}
($\approx 130$~AU) from \epseri.
We detected four of these objects (numbered 3, 5, 6
and 10) in at least 2 IRAC images. We could not detect sources 1, 2,
4, 7, 8 and 9, which are below the PSF-subtraction residual noise in
their position. We have however
detected one source that was missed in the Keck search (inside the
square box number 11). A careful examination of the Keck
plates (Figure~2 in \citealt{macintosh03}) indeed shows the presence
of a faint source (below the Keck observations detection limit) in the
location of source 11. Sources 4, 6 and 11 have also been
detected by \citet{proffitt04}: they have an extended sources
appearance and are probably background galaxies.

Table~\ref{tab-macintosh} summarizes the properties of the
\citet{macintosh03} sources. The ``Keck Coordinates'' columns indicate
the position of each source at the time of the \citet{macintosh03}
first epoch (December 2001). Since Table~3 in \citet{macintosh03}
does not provide absolute coordinates for the sources, but only
offsets from \epseri, we had to derive their actual R.A and Dec by
estimating the position of \epseri{} in December 2001 from the ICRS
2000 \citep{perryman97} coordinates and proper motion. Given that
\citet{macintosh03} had to keep \epseri{} outside their field of view
to limit PSF artifacts, the accuracy of the December 2001 coordinates
is limited to 0.2\arcsec, which is the precision with which
\citet{macintosh03} can establish the position of \epseri{} relative to
their images. We then estimated the coordinates of each
source at the time of the first IRAC epoch (January 2004) by
extrapolating the motion measured by \citet{macintosh03} between
December 2001 and their second epoch in August 2002. Note that for
some of the sources, either because they were used as position
reference, or because the second epoch was missing (source 3), the
predicted 2004 position cannot be derived. Given the high accuracy
quoted by \citet{macintosh03} in the relative shift measurements, the
uncertainty of the predicted coordinates in the IRAC images is still
dominated by the uncertainty in their 2001 absolute coordinate, of
0.2\arcsec. 

Finally we determined the position of all sources detected within
the Keck field of view by measuring their photo-center in the IRAC
4.5~\micron{} images. The uncertainty
in the measurement of the sources centroid is limited by the presence
of PSF subtraction artifacts and second epoch source aliasing, but is
consistent, between our two epochs, with an accuracy of
0.2\arcsec. Within this limit, we didn't observe any shift in the
source coordinates between the two IRAC epochs. The real uncertainty
in the IRAC coordinates, however, is much larger than 0.2\arcsec, due
to systematic errors in the pointing of the spacecraft. These errors
are normally corrected by a pointing refinement procedure, included in
the Spitzer Space Center data reduction pipeline, that uses
2MASS stars \citep{cutri03} to find an absolute reference frame for
the IRAC images within the 0.15\arcsec{} 2MASS accuracy. In the case
of \epseri, however, this pointing refinement was not done because
there isn't any 2MASS source detected in the vicinity (because of the
\epseri{} glare saturating the 2MASS detectors). The headers of the
IRAC BCD frames show a pointing accuracy error of $\sim 1$\arcsec{} at
the time of our observations.

By comparing the coordinates of all sources detected in both Keck
and IRAC images, we have measured the shift between December 2001 and
January 2004. The values we obtained, listed in Table~\ref{tab-macintosh}
are rather large, and appear to suffer a systematic shift, possibly due
to the limitations in the Spitzer pointing accuracy described
above. We have tried to correct this systematic shift by subtracting
the average value of the the measured shift, of 0.13\arcsec{} in
R.A. and 1.30\arcsec{} in Dec. We estimate the uncertainty of the
final shifts to be of the order of $\sim 1$\arcsec. A significant
fraction of this uncertainty may be due to the different centroid that
partially resolved extragalactic sources, if extended or multiple,
may have in the K and 4.5~\micron{} bands. Within this error,
none of the sources show a measurable proper motion, and the corrected
shifts listed in Table~\ref{tab-macintosh} are all well below the
expected \epseri{} proper motion between the Keck and IRAC
observations, of more than 2\arcsec{} in R.A. We also did not find any
significant shift between the coordinates of sources 4, 6 and 11 as
measured in the IRAC 4.5~\micron{} images, and in the STIS/HST
\citet{proffitt04} maps. 

This confirms that none of the sources detected in the common field of
view with the Keck and \citet{proffitt04} observations is a low mass
companion of \epseri.


\section{Conclusions}\label{sec-conclude}

As the number of exoplanets detected with radial velocity or
eclipsing techniques, since the initial detection by \citet{mayor95},
continue to grow, the payoff of directly detecting
their thermal emissions cannot be underestimated. As of today, thermal
radiation has been directly detected from only three planets,
all of them eclipsing ``hot Jupiters'' whose light has been detected
with Spitzer: TrES-1 with IRAC \citep{charbonneau05} at 4.5 and
8.0~\micron{}, HD209458b \citep{deming05} with MIPS
24~\micron{} and HD189733b with IRS \citep{deming06}.
These three detections allowed for the first time a
comparison between theoretical models of hot Jupiters with real
data. However, no direct detection of gas giants  orbiting a nearby
parent star at a distance comparable to our own Jupiter and Saturn has
been achieved so far, as all searches conducted from the ground
\citep{macintosh03} and from space (\citealt{proffitt04} and this one)
have returned negative results. Models of giant planets orbiting 
systems analogous to the Solar System are thus still untested.

While future space missions, such as the James Webb Space telescope
and the Terrestrial Planet Finder are being designed with the goal of
direct exoplanetary detection, this work shows that the Spitzer Space
telescope with the IRAC camera already has the capability of 
detecting young gas giants with a mass as low as  a few \mj{} if they
are orbiting their parent star at a distance comparable to our own
Solar System Kuiper belt. Planets ejected on very elongated orbits
during the initial phases of planetary system formation can
be detected with Spitzer/IRAC down to masses of 1~\mj. Our
observations allow to set a limit of $\sim 2$~\mj{} at the distance of
$\sim 60$~AU ($\sim 18$\arcsec) from the star, and $\la 1$\mj{}
outside the debris disk ring ($\sim 35$\arcsec $\simeq 112$~AU).

Comparison with \citet{macintosh03} confirms that within our
search space we are able to detect all the sources found with the most
sensitive ground based search outside a radius of $\sim 130$~AU and,
in addition, one below the Keck sensitivity. Comparison
with the position of the common sources between IRAC and Keck
observations confirms that none of the detections match the proper
motion required to be a physical companion of \epseri, validating and
extending \citet{macintosh03} results.

Based on the colors of the detected sources, we did not find any
strong companion candidate. However, for a number of sources, listed
in Table~\ref{tab-dropout}, we are missing a detection in the
3.6~\micron{} IRAC band. Without a good [3.6]-[4.5] color we cannot
determine the nature of these sources, and a second epoch is necessary
to measure their proper motion. Given the positional accuracy in the
PSF-subtracted IRAC frames of $\sim 1$\arcsec, such a measurement will
be feasible after a time lag of about 2 years from the current
observations.


\acknowledgments

We would like to thank Brian Patten for providing the IRAC
photometry of the MLT dwarfs we used as templates for our $k$NN
method, Robert Gutermuth for providing the PHOTVIS software, Mayly
C. Sanchez for helping to adapt the $k$NN method to this application,
Jiasheng Huang for kindly providing unpublished data used to determine
the statistics and IRAC colors of extragalactic sources
and Michael Schuster for a careful reading of this manuscript. 

This work is based on observations made with the Spitzer Space
Telescope, which is operated by the Jet Propulsion Laboratory,
California Institute of Technology under NASA contract 1407. Support
for the IRAC instrument was provided by NASA under contract number
1256790 issued by JPL. This publication makes use of data products
from the Two Micron All Sky Survey, which is a joint project of the
University of Massachusetts and the Infrared Processing and Analysis
Center/California Institute of Technology, funded by the National
Aeronautics and Space Administration and the National Science
Foundation. 



\clearpage








\begin{deluxetable}{lcccc}
\tablecaption{IRAC Photometric Calibration\tablenotemark{a}\label{tab-cal}}
\tablewidth{1\textwidth}
\tablehead{ & \multicolumn{4}{c}{IRAC band}\\
\colhead{Item} &
\colhead{3.6~\micron} &
\colhead{4.5~\micron} &
\colhead{5.8~\micron} &
\colhead{8.0~\micron} }
\startdata
Isophotal $\lambda$ [\micron] & 3.458  & 4.492  & 5.661  & 7.870  \\
FLUXCONV [(MJy/sr)/(DN/s)]     & 0.1125 & 0.1375 & 0.5913 & 0.2008 \\
Saturation limits [DN]         & 30,000 & 35,000 & 45,000 & 50,000 \\
Saturation limits [MJy/sr]     & 320    & 460    & 2560   & 970    \\
Infinite aperture corrections  & 0.94   & 0.94   & 0.63   & 0.69   \\
2.4\arcsec{} aperture corrections\tablenotemark{b}
                               & 1.213  & 1.234  & 1.379  & 1.584  \\
Zero point magnitudes\tablenotemark{c}
                               & 19.46  & 18.97  & 18.38  & 17.56  \\
$F_\nu$(Vega) [Jy]             & 277.5  & 179.5  & 116.5  & 63.1   \\
\enddata
\tablenotetext{a}{Based on \citet{irac04}}
\tablenotetext{b}{For sky annulus with 2.4\arcsec{} and 7.2\arcsec{}
  inner and outer radii}
\tablenotetext{c}{For pixel size 0.4\arcsec/pix, include 2.4\arcsec{}
  aperture correction}
\end{deluxetable}
\clearpage

\begin{deluxetable}{cccccc}
\tablecaption{\epseri{} IRAC magnitudes and fluxes\label{tab-mags}} 
\tablewidth{1\textwidth}
\tablehead{ & &
\multicolumn{2}{c}{January 9, 2004} &
\multicolumn{2}{c}{February 17, 2004} \\
\colhead{IRAC band} & 
\colhead{$\lambda_0$ [\micron]\tablenotemark{a}} &
\colhead{mag\tablenotemark{b}} &
\colhead{Flux [Jy]\tablenotemark{c}} &
\colhead{mag\tablenotemark{b}} &
\colhead{Flux [Jy]\tablenotemark{c}}}
\startdata
[3.6] & 3.548 & 1.59 & 63.9$\pm$0.3 & 1.59 & 64.1$\pm$0.4 \\

[4.5] & 4.492 & 1.66 & 39.1$\pm$0.2 & 1.65 & 39.2$\pm$0.2 \\

[5.8] & 5.661 & 1.61 & 26.5$\pm$0.2 & 1.64 & 25.8$\pm$0.2 \\

[8.0] & 7.870 & 1.60 & 14.5$\pm$0.1 & 1.59 & 14.6$\pm$0.1 \\
\enddata
\tablenotetext{a}{IRAC nominal wavelengths from the \citet{irac04}}
\tablenotetext{b}{Magnitudes accurate to $\pm$0.01~mag}
\tablenotetext{c}{Using IRAC Vega fluxes from the \citet{irac04}}
\end{deluxetable}
\clearpage

\begin{deluxetable}{rcccccc}
\tablecaption{Sources with marginal or no detection at 3.6~\micron{}
  within 600 A.U. from \epseri\label{tab-dropout}} 
\tablewidth{1\textwidth}
\tabletypesize{\footnotesize}
\tablehead{ \colhead{\#} & \colhead{RA} & \colhead{Dec} & \colhead{[3.6]} &
            \colhead{[4.5]} & \colhead{[3.6]$-$[4.5]} & \colhead{dist (AU)} }
\startdata
 1 & 53.189347 & -9.444469 & $18.0\pm0.6$ & $17.29\pm0.19$ & $0.3\pm0.6$ & 512\\
 2 & 53.194310 & -9.465338 & $19.2\pm0.7$ & $18.65\pm0.07$ & $0.5\pm0.7$ & 437\\
 3 & 53.194881 & -9.465428 & $19.0\pm0.7$ & $18.52\pm0.05$ & $0.5\pm0.7$ & 430\\
 4 & 53.201029 & -9.469290 & $19.7\pm0.7$ & $19.59\pm0.15$ & $0.1\pm0.7$ & 373\\
 5\tablenotemark{a}
   & 53.205047 & -9.452265 & $> 20.2$     & $17.30\pm0.17$ & $> 2.9$     & 312\\
 6\tablenotemark{b}
   & 53.205817 & -9.486791 & $> 20.5$     & $19.14\pm0.10$ & $> 1.4$     & 443\\
 7 & 53.206656 & -9.489741 & $19.6\pm0.6$ & $19.32\pm0.11$ & $0.3\pm0.6$ & 464\\
 8 & 53.208558 & -9.492060 & $19.7\pm0.6$ & $18.96\pm0.06$ & $0.7\pm0.6$ & 472\\
 9 & 53.213247 & -9.454328 & $19.0\pm0.6$ & $18.66\pm0.07$ & $0.3\pm0.6$ & 214\\
10 & 53.213330 & -9.447145 & $19.5\pm0.6$ & $18.43\pm0.07$ & $1.1\pm0.6$ & 245\\
11 & 53.213868 & -9.414921 & $17.8\pm0.5$ & $17.70\pm0.08$ & $0.1\pm0.5$ & 543\\
12 & 53.219017 & -9.474964 & $19.2\pm0.7$ & $18.42\pm0.07$ & $0.8\pm0.7$ & 240\\
13 & 53.219744 & -9.481340 & $19.4\pm0.8$ & $17.57\pm0.08$ & $1.8\pm0.8$ & 299\\
14 & 53.221378 & -9.448212 & $18.7\pm0.8$ & $18.38\pm0.11$ & $0.3\pm0.8$ & 164\\
15 & 53.223314 & -9.488778 & $19.7\pm0.6$ & $18.50\pm0.07$ & $1.2\pm0.6$ & 366\\
16 & 53.227599 & -9.478849 & $19.0\pm0.6$ & $18.28\pm0.05$ & $0.3\pm0.6$ & 242\\
17 & 53.229451 & -9.440029 & $18.9\pm0.6$ & $17.85\pm0.05$ & $1.0\pm0.6$ & 213\\
18 & 53.234574 & -9.413988 & $18.6\pm0.4$ & $17.76\pm0.07$ & $0.8\pm0.4$ & 516\\
19 & 53.236470 & -9.497356 & $20.7\pm1.9$ & $19.13\pm0.09$ & $1.6\pm1.9$ & 458\\
20 & 53.237040 & -9.504618 & $19.3\pm0.7$ & $18.44\pm0.06$ & $0.9\pm0.7$ & 542\\
21 & 53.238661 & -9.440854 & $19.4\pm0.9$ & $18.46\pm0.06$ & $0.9\pm0.9$ & 220\\
22 & 53.240732 & -9.441868 & $19.1\pm0.9$ & $18.63\pm0.09$ & $0.5\pm0.9$ & 220\\
23 & 53.249281 & -9.498276 & $19.1\pm0.9$ & $18.60\pm0.08$ & $0.5\pm0.9$ & 509\\
24 & 53.253290 & -9.495511 & $17.7\pm0.3$ & $17.54\pm0.03$ & $0.2\pm0.3$ & 502\\
25 & 53.258599 & -9.484475 & $19.3\pm0.7$ & $18.48\pm0.05$ & $0.8\pm0.7$ & 440\\
26 & 53.259360 & -9.471248 & $19.1\pm0.6$ & $18.97\pm0.09$ & $0.1\pm0.6$ & 359\\
27 & 53.268980 & -9.431036 & $18.4\pm0.4$ & $17.70\pm0.03$ & $0.7\pm0.4$ & 541\\
28\tablenotemark{c}
   & 53.269965 & -9.487714 & $> 20.2$     & $17.64\pm0.16$ & $> 2.6$     & 565\\
29 & 53.272692 & -9.430506 & $18.7\pm0.5$ & $18.34\pm0.07$ & $0.4\pm0.5$ & 580\\
\enddata
\tablenotetext{a}{Multiple or extended sources detected in all
  4.5~\micron{} images.}
\tablenotetext{b}{Point source (possibly double) detected in second
  epoch 4.5~\micron{} images.} 
\tablenotetext{c}{Point source (possibly multiple) detected in second
  epoch 4.5~\micron{} images.}
\end{deluxetable}
\clearpage

\begin{deluxetable}{cccccccccccccccc}
\tablecaption{Proper motion and photometry of \citet{macintosh03}
  objects\label{tab-macintosh}} 
\tablewidth{1\textheight}
\rotate
\tabletypesize{\footnotesize}
\tablehead{\colhead{\#} &
           \multicolumn{2}{c}{Keck Coordinates\tablenotemark{a}} &
           \multicolumn{2}{c}{IRAC Coordinates} &
           \multicolumn{2}{c}{Predicted\tablenotemark{b}} &
           \multicolumn{2}{c}{Measured\tablenotemark{b,c}} &
           \multicolumn{2}{c}{Corrected\tablenotemark{b,c,d}} &
           \multicolumn{5}{c}{Photometry\tablenotemark{e}} \\
           \colhead{} &
           \colhead{R.A.} & \colhead{Dec} &
           \colhead{R.A.} & \colhead{Dec} &
           \colhead{$\Delta \alpha$} & \colhead{$\Delta \delta$} &
           \colhead{$\Delta \alpha$} & \colhead{$\Delta \delta$} &
           \colhead{$\Delta \alpha$} & \colhead{$\Delta \delta$} &
           \colhead{[K]} & \colhead{[3.6]} & \colhead{[4.5]} &
                           \colhead{[5.8]} & \colhead{[8.0]}
}
\startdata
 \epseri & 53.23216 & -9.45825 & 53.23159 & -9.45824 &
           2.05     & -0.04    &          &          &          &          &
                    &          &          &          &          \\
1        & 53.22949 & -9.45431 & ---      & ---      &
           ---      & ---      & ---      & ---      & ---      & ---      &
           17.3     & ---      & ---      & ---      & ---      \\
2        & 53.23341 & -9.45353 & ---      & ---      &
           ---      & ---      & ---      & ---      & ---      & ---      &
           17.3     & ---      & ---      & ---      & ---      \\
3        & 53.23099 & -9.44600 & 53.23085 & -9.44634 &
           ---      & ---      & 0.50     & 1.22     & 0.37     & -0.08    &
           16.3     & 15.4     & 15.4     & 15.5     & 15.4     \\
4\tablenotemark{f}        & 53.23960 & -9.45539 & ---      & ---      &
           -0.03    & 0.01     & ---      & ---      & ---      & ---      &
           19.4     & ---      & ---      & ---      & ---      \\
5        & 53.24291 & -9.45422 & 53.24316 & -9.45455 &
           0.05     & 0.28     & -0.90    & 1.19     & -1.03    & -0.11    &
           20.7     & 17.4     & 16.5     & 15.1     & 15.7     \\
6\tablenotemark{f}        & 53.24219 & -9.46369 & 53.24212 & -9.46390 &
           0.09     & -0.03    & 0.25     & 0.76     & 0.12     & -0.54    &
           20.2     & 17.6     & 17.1     & ---      & ---      \\
7        & 53.22852 & -9.45369 & ---      & ---      &
           0.05     & 0.16     & ---      & ---      & ---      & ---      &
           20.3     & ---      & ---      & ---      & ---      \\
8        & 53.22343 & -9.45447 & ---      & ---      &
           0.08     & 0.00     & ---      & ---      & ---      & ---      &
           20.1     & ---      & ---      & ---      & ---      \\
9        & 53.23658 & -9.46461 & ---      & ---      &
           -0.10    & 0.06     & ---      & ---      & ---      & ---      &
           20.8     & ---      & ---      & ---      & ---      \\
10       & 53.22722 & -9.46772 & 53.22703 & -9.46828 &
           0.06     & 0.10     & 0.68     & 2.01     & 0.55     & 0.71     &
           19.3     & 16.7     & 16.2     & ---      & 15.7     \\
11\tablenotemark{f}       & ---      & ---      & 53.24301 & -9.45902 &
           ---      & ---      & ---      & ---      & ---      & ---      &
           ---      & ---      & 16.5     & ---      & 15.3     \\
\enddata
\tablenotetext{a}{Positions calculated for Keck first epoch, Dec 1,
  2001}
\tablenotetext{b}{Shifts expressed in second of arc}
\tablenotetext{c}{Shifts statistical uncertainty is 0.2\arcsec}
\tablenotetext{d}{Shifts corrected by subtracting the average shift of
  0.13\arcsec{} in R.A. and 1.30\arcsec{} in Dec}
\tablenotetext{e}{Average statistical error in IRAC photometry is 0.2 mag}
\tablenotetext{f}{Detected in \citet{proffitt04} as extended structures}
\end{deluxetable}
\clearpage




\clearpage

\clearpage

\clearpage

\clearpage

\clearpage

\clearpage

\clearpage

\clearpage

\clearpage

\end{document}